\documentclass[aps,pre,reprint,superscriptaddress,longbibliography]{revtex4-2}

\usepackage{amsmath}
\usepackage{amssymb}
\usepackage{bm}
\usepackage{physics}
\usepackage{graphicx}
\usepackage{xcolor}
\usepackage{lineno} 
\usepackage[
    colorlinks=true,
    linkcolor=blue,
    citecolor=blue,
    urlcolor=blue
]{hyperref}
\begin{document}

\title{Synchronization induced by stochastic subsystem resetting in complex networks}

\author{Kevin Paramero-Arias${}^1$, Alejandro P. Riascos}
\address{Departamento de Física, Universidad Nacional de Colombia, Bogotá, Colombia}

\begin{abstract}
We study synchronization in network-coupled oscillators subject to stochastic resetting of a subset of nodes. We show that resetting can induce collective coherence in oscillators that are never directly reset, through the interplay of network structure, frequency heterogeneity, and the organization of the reset subsystem. In the infinite-resetting-rate limit, the reset nodes become effectively pinned to a reference phase, and linearization near synchronization maps the dynamics of the remaining subsystem onto a grounded-Laplacian problem. This formulation yields an analytical relation between the synchronization level and the reset parameters, providing an estimate of the coupling strength required to achieve a prescribed degree of coherence. Beyond the linear regime, we develop a mean-field description in which the reset subsystem acts as an effective ordering field for the remaining oscillators. Numerical simulations across different network topologies confirm these predictions and show that synchronization is enhanced by increasing both the reset fraction and network connectivity. In heterogeneous networks, the placement of reset nodes can strongly influence the synchronization response. For frequency distributions with nonzero mean, resetting competes with the intrinsic collective rotation and can drive the system from a rotating synchronized state to a phase-locked state. These results establish subsystem resetting as a mechanism for inducing and controlling collective synchronization in complex networks.
\end{abstract}

\maketitle
\section{Introduction}
Synchronization is a fundamental collective phenomenon observed in a
wide range of physical, biological, and technological systems,
including coupled oscillators, neuronal populations, power grids,
and communication networks
\cite{BOCCALETTI2008,strogatzbook,Balanov2009,J_Kurths_PhysRep2023}.
Its ubiquity is illustrated by phenomena ranging from the synchronized
flashing of fireflies \cite{PikoBook,Peleg_2021} and collective
clapping in large crowds \cite{Neda_2000} to synchronized dynamics
in arrays of Josephson junctions
\cite{Wiesenfeld_PRL1996}. A paradigmatic framework for investigating
the emergence of synchronization is the Kuramoto model
\cite{yoshiki_k}, which describes a population of phase oscillators
with heterogeneous natural frequencies coupled through interactions
that promote phase alignment
\cite{PikoBook,strogatzbook,Acebron_RevMod_Phys2005,J_Kurths_PhysRep2023}.
Despite its simplicity, this model provides a powerful framework for
understanding the transition from incoherence to collective
synchronization \cite{Strogatz_PhysD2000}. In many real-world systems, however, interactions are not global but
are constrained by an underlying network
\cite{Restrepo_PRE_2005,DORFLER2014,Rodrigues2016}. The networked
Kuramoto model accounts for this constraint by coupling oscillators
according to the network's connectivity, making topology an
intrinsic determinant of collective dynamics
\cite{Arenas2008,Rodrigues2016}. Consequently, synchronization
depends not only on the coupling strength and the distribution of
natural frequencies, but also on structural properties such as
connectivity, degree heterogeneity, and global network organization
\cite{Restrepo_PRE_2005,Gomez_G_PRE2007,Gomez_G_PRL2007,BOCCALETTI2008,Arenas2008,Gomez_G_PRL2011,Rodrigues2016}.
Recent studies have further demonstrated how network structure and
frequency heterogeneity shape synchronization transitions, including
abrupt transitions triggered by link perturbations and the effects
of frequency shuffling
\cite{ArolaFernandez2022,Manaoj_PRE_2024,Pruser_PRL_2024}.
Comparisons of synchronization processes have also highlighted the
role of network topology in shaping collective behavior
\cite{Riascos_PRE_2024,Polo-Gonzalez_2025,ErasoHernandez2026Antifragility}.
\\[2mm]
On the other hand, stochastic resetting provides a versatile mechanism for modifying nonequilibrium dynamics by intermittently interrupting a process and reinstating it according to a prescribed protocol. The seminal study of diffusion with Poissonian resetting showed that such interruptions can generate nontrivial stationary states and modify first-passage properties \cite{Evans_PRL_2011}. Subsequent work extended resetting to time-dependent and non-Poissonian protocols \cite{Pal_2016,Nagar_PRE_2016}, external potentials and bounded domains \cite{Pal_PRE_2015,Pal_PRE_2019}, and spatially distributed reset positions \cite{Salgado_PRE_2025}. Resetting has also been investigated in systems with fluctuating targets and interacting particles, revealing its interplay with intrinsic dynamical processes \cite{Mercado_JPhysA_2021,Nagar_2023}. More recently, resetting has been shown to generate emergent correlations in noninteracting Brownian gases, affecting extreme-value statistics and interparticle spacings, while batch resetting provides a mechanism to tune their strength \cite{BiroliPRL_2023,de_Mauro_2026}. These findings exemplify dynamically emergent correlations in the absence of direct interactions \cite{majumdar2026dynamically} and further establish stochastic resetting as a versatile framework for modifying stationary, first-passage, and collective dynamics in nonequilibrium systems \cite{Evans_2020,Nagar_2023}. 
\\[2mm]
When the underlying dynamics takes place on a network, resetting becomes coupled to the topology and can modify transport, exploration, and first-passage processes. Random walks with resetting have been studied on arbitrary networks, where spectral approaches show that resetting can either enhance or hinder search efficiency depending on the network structure and target \cite{Riascos_PRE_2020,Multiple_Reset_PRE_2021,Riascos_2022}. These results have been extended to fractal and scale-free networks and to biased random walks \cite{Peng_PRE_2023,Sun_Entropy_2024,Li_Chaos_2025,Li_PRE_2025}. More general protocols include centrality-based and non-Markovian resetting \cite{Zelenkovski_Entropy_2023,Michelitsch_Chaos_2025}, while resetting the memory of self-avoiding walkers provides another mechanism for controlling network exploration \cite{Colombani_2023}. Recent studies have also considered multiple walkers and simultaneous resetting, as well as the effect of resetting on extreme events in networked transport \cite{Rubio-Gomez_PRE_2025,Suarez_PRE_2026,Sun_Entropy_2025}. These results highlight the interplay between stochastic resetting and network topology in shaping dynamical processes. 
\\[2mm]
In the context of interacting oscillator systems, resetting can strongly affect collective behavior and has been explored in several settings, including globally reset populations \cite{Sarkar2022}, density-based formulations of interacting systems \cite{Bressloff2024}, and oscillator populations coupled through an external medium \cite{Bressloff2025}. Extensions have also considered intrinsic noise and general frequency distributions \cite{Acharya2025}, as well as adaptive coupling in partially reset populations \cite{Li2025}, while recent work on phase resetting has demonstrated that large phase shifts in coupled oscillator populations can be achieved more efficiently through transient desynchronization followed by resynchronization at the desired phase, highlighting the role of collective synchronization dynamics in resetting processes \cite{Wilson2026}. The Kuramoto model provides a natural framework for studying these effects. It describes globally coupled phase oscillators with distributed natural frequencies, exhibiting a transition from incoherence to synchronization as the coupling strength exceeds a critical value. Introducing stochastic resetting adds a second time scale and can induce synchronization even in parameter regimes where the unreset system remains incoherent \cite{Sarkar2022}. This naturally raises the question of whether synchronization can be controlled by resetting only a fraction of the oscillators. This subsystem-resetting problem was introduced in the globally coupled Kuramoto model \cite{Majumder_PRE_2024,Majumder_PRE2026}, where a subset of oscillators is repeatedly reset to a synchronized state while the remainder evolves according to the Kuramoto dynamics. Resetting can induce synchronization in the non-reset population, and in the limit of infinitely fast resetting the reset population acts as a coherent forcing. In this limit, the resulting dynamics can be analyzed using the Ott--Antonsen approach \cite{Ott_Chaos2008}, with analytical results agreeing with numerical simulations for representative frequency distributions \cite{Majumder_PRE_2024,Majumder_PRE2026}.
\\[2mm]
The extension of this problem to complex networks is nontrivial because the fully connected geometry suppresses the role of the spatial organization of the reset subsystem. In an arbitrary network, reset oscillators do not couple equally to all other oscillators, and their influence on the non-reset subsystem depends on the topology of the network and on the particular set of nodes that is reset. Consequently, two reset configurations with the same number of reset nodes can produce different synchronization responses. In the limit of strong synchronization and rapid resetting, the reset subsystem can be viewed as a set of nodes pinned to a reference phase. The resulting dynamics of the non-reset subsystem is therefore naturally related to network-pinning problems and to the spectral properties of a grounded Laplacian \cite{Pirani2014,Pirani2016,Xia2017,He2019,Ziang_2026}. This provides a route to quantify how the topology and the placement of the reset nodes determine the synchronization induced in the non-reset subsystem. Beyond the strongly synchronized regime, a complementary collective description is needed to characterize the nonlinear response of the network to the resetting-induced coherence.
\\[2mm]
In this paper, we extend the study of subsystem resetting to Kuramoto oscillators coupled through complex networks. We focus on the limit of infinitely fast resetting, where the reset oscillators become effectively pinned to a reference phase, and develop a linear theory for the dynamics of the non-reset subsystem. In the strongly synchronized regime, this leads naturally to a grounded-Laplacian description, which relates the stationary synchronization to the network topology, the distribution of natural frequencies, and the placement of the reset nodes. We further develop a mean-field theory for dense random networks that describes the system beyond the linear regime and interprets the reset subsystem as an effective ordering field. The analytical predictions are tested through numerical simulations on several network topologies, including Erd\H{o}s--R\'enyi, Watts--Strogatz, and Barab\'asi--Albert networks. We show that the synchronization induced by resetting depends not only on the fraction of reset nodes and the coupling strength, but also on their structural location in the network. We also consider frequency distributions with nonzero mean and characterize the competition between resetting-induced synchronization and collective rotation. Our results establish a framework for understanding and controlling synchronization through localized resetting in complex networks.
\section{General Theory}

\subsection{Kuramoto dynamics with subsystem resetting}

We consider a system of $N$ coupled phase oscillators interacting through an undirected connected network. The structure of the network is characterized by the adjacency matrix $\mathbf{A}$, whose elements $A_{ij}$ take the value $A_{ij}=1$ when nodes $i$ and $j$ are connected, and $A_{ij}=0$ otherwise, for $i,j=1,2,\ldots,N$. Since the network is undirected, the adjacency matrix is symmetric, namely, $A_{ij}=A_{ji}$. Throughout this work, we consider connected networks, implying that every pair of nodes is linked through at least one path. The degree of node $j$ is defined as $k_j=\sum_{l=1}^{N} A_{jl}$ which gives the total number of connections associated with node $j$.
\\[2mm]
At every node of the network there is a phase oscillator characterized by a time-dependent phase $\theta_i(t)\in[0,2\pi)$. The dynamics of the oscillators is governed by a generalized Kuramoto model on networks \cite{yoshiki_k,Rodrigues2016}
\begin{equation}
	\frac{d\theta_i}{dt}
	=
	\omega_i
	+
	\frac{K}{N}
	\sum_{j=1}^{N}
	A_{ij}
	\sin(\theta_j-\theta_i),
\end{equation}
where $\omega_i$ denotes the natural frequency of oscillator $i$, and $K>0$ represents the coupling strength between connected oscillators. The adjacency matrix determines the interaction pattern of the system, such that oscillator $i$ interacts only with oscillators connected to it through the network. The natural frequencies $\omega_i$ are quenched random variables drawn from a probability density $g(\omega)$.
\\[2mm]
The oscillators are separated into two subsets. A fraction of nodes belongs to the reset subsystem, while the remaining oscillators constitute the non-reset subsystem. We denote the set of reset nodes by $\mathcal{R}$ and the set of non-reset nodes by $\mathcal{M}$, with size $|\mathcal{M}| = M$. Accordingly, the reset subsystem contains $N-M$ oscillators. The subsets satisfy
\begin{equation}
	\mathcal{R}\cup \mathcal{M}
	=
	\{1,2,\ldots,N\},
	\qquad
	\mathcal{R}\cap \mathcal{M}
	=
	\varnothing.
\end{equation}
The dynamics evolves continuously according to the Kuramoto equations, interrupted by stochastic resetting events acting exclusively on the oscillators belonging to the reset subsystem. In this framework, the oscillators in $\mathcal{M}$ evolve solely under the bare dynamics, whereas the oscillators in $\mathcal{R}$ undergo alternating sequences of free evolution and resetting events.
\\[2mm]
The resetting events occur according to a Poisson process with rate $\lambda$. Therefore, during an infinitesimal time interval $dt$, a resetting event takes place with probability $\lambda dt$, while with probability $1-\lambda dt$ all oscillators evolve according to the bare Kuramoto dynamics. At every resetting event, the phases of the oscillators belonging to the reset subsystem are instantaneously set to a prescribed reference configuration. In the present work, we consider the synchronized resetting condition
\begin{equation}\label{Equ_Theta_R}
	\theta_j \rightarrow 0,
	\qquad
	j\in \mathcal{R}.
\end{equation}
Consequently, the resetting mechanism continuously injects local phase coherence into the system through the subset $\mathcal{R}$. By contrast, the oscillators belonging to the non-reset subsystem are never directly reset, and their collective behavior emerges exclusively from the interplay between the network interactions and the coupling with the reset subsystem. The waiting times $\tau$ between consecutive resetting events are independent random variables distributed according to
\begin{equation}
	P(\tau)
	=
	\lambda e^{-\lambda \tau},
	\qquad
	\tau \geq 0.
\end{equation}
To characterize synchronization in the complete system, we use the order parameter \cite{Arenas2008, Rodrigues2016, J_Kurths_PhysRep2023}
\begin{equation}\label{r_psi_def}
	re^{\mathbf{i}\psi}
	=
	\frac{1}{N}
	\sum_{j=1}^{N}
	e^{\mathbf{i}\theta_j},
\end{equation}
where $\mathbf{i}\equiv \sqrt{-1}$. In Eq. (\ref{r_psi_def}), $0\leq r\leq 1$ quantifies the degree of global phase coherence, while $\psi$ represents the average phase of the oscillators.
\\[2mm]
Additionally, it is useful to define order parameters associated separately with the reset and non-reset subsystems. For the non-reset subsystem given by \cite{Majumder_PRE_2024}
\begin{equation}\label{Order_p_non-reset_M}
	r_{\mathcal{M}}
	e^{\mathbf{i}\psi_{\mathcal{M}}}
	=
	\frac{1}{M}
	\sum_{j\in\mathcal{M}}
	e^{\mathbf{i}\theta_j},
\end{equation}
whereas for the reset subsystem \cite{Majumder_PRE_2024}
\begin{equation}
	r_{\mathcal{R}}
	e^{\mathbf{i}\psi_{\mathcal{R}}}
	=
	\frac{1}{N-M}
	\sum_{j\in\mathcal{R}}
	e^{\mathbf{i}\theta_j}.
\end{equation}
The quantities $r_{\mathcal{M}}$ and $r_{\mathcal{R}}$ allow us to characterize separately the synchronization properties of both subsystems and to quantify how resetting a fraction of oscillators affects the emergence of coherence in oscillators that are not directly reset. Our main goal is to analyze how the interplay between the network topology, the resetting rate $\lambda$, and the size of the reset subsystem determines the collective synchronization properties of the system.
\subsection{Infinite resetting-rate limit}

In the limit of infinite resetting rate, namely $\lambda\to\infty$, the oscillators belonging to the reset subsystem remain permanently pinned to the resetting phase. Without loss of generality, the reset phase may be chosen equal to zero as in Eq. (\ref{Equ_Theta_R}). As a consequence, the reset subsystem becomes dynamically frozen and acts continuously on the oscillators belonging to the non-reset subsystem. In this limit, only the oscillators in $\mathcal{M}$ evolve dynamically, while the reset nodes provide an effective forcing contribution mediated by the network connectivity.
\\[2mm]
For a non-reset oscillator $i\in\mathcal{M}$, the dynamics takes the form
\begin{equation}
	\dot{\theta}_i
	=
	\omega_i
	+
	\frac{K}{N}
	\sum_{j\in\mathcal{M}}
	A_{ij}\sin(\theta_j-\theta_i)
	+
	\frac{K}{N}
	\sum_{j\in\mathcal{R}}
	A_{ij}\sin(\theta_j-\theta_i).
\end{equation}
Using the condition $\theta_j=0$ for all $j\in\mathcal{R}$, we obtain
\begin{equation}\label{Kuramoto_OD_reset}
	\dot{\theta}_i
	=
	\omega_i
	+
	\frac{K}{N}
	\sum_{j\in\mathcal{M}}
	A_{ij}\sin(\theta_j-\theta_i)
	-
	\frac{K}{N}
	\sum_{j\in\mathcal{R}}
	A_{ij}\sin\theta_i .
\end{equation}
To characterize the local influence of the reset subsystem on node $i$, it is convenient to define the quantity
\begin{equation}
	q_i
	=
	\sum_{j\in\mathcal{R}}
	A_{ij},
\end{equation}
which counts the number of reset neighbors connected to node $i$. Using this definition, the dynamics of the non-reset oscillators can be rewritten as
\begin{equation}\label{KM_inf_lambda}
	\dot{\theta}_i
	=
	\omega_i
	+
	\frac{K}{N}
	\sum_{j\in\mathcal{M}}
	A_{ij}\sin(\theta_j-\theta_i)
	-
	\frac{K}{N}
	q_i\sin\theta_i .
\end{equation}
Equation (\ref{KM_inf_lambda}) reveals that, in the infinite resetting-rate limit, the reset subsystem induces an effective local pinning field acting on the non-reset oscillators. The strength of this contribution is not uniform across the network and depends explicitly on the number of reset neighbors connected to each node. Consequently, the synchronization properties of the non-reset subsystem are determined not only by the coupling strength $K$, but also by the interplay between the network topology and the distribution of reset nodes encoded in the quantities $q_i$.

\subsection{Linearized dynamics and grounded Laplacian}
\label{Sub_Sec_linear_Lg}
We now focus on the strongly synchronized regime, in which the phases of the non-reset oscillators remain close to the reset phase. In this case, the phase differences between connected oscillators are small, allowing a linear approximation of the dynamics. Specifically, we assume
\begin{equation}
	|\theta_i| \ll 1,
\end{equation}
for all $i\in\mathcal{M}$. Under this condition, the interaction terms may be expanded to leading order as
\begin{equation}
	\sin(\theta_j-\theta_i)
	\approx
	\theta_j-\theta_i,
	\qquad
	\sin\theta_i
	\approx
	\theta_i .
\end{equation}
Substituting these approximations into the dynamics obtained in the infinite resetting-rate limit yields
\begin{equation}\label{KM_linear_non-reset}
	\dot{\theta}_i
	\approx
	\omega_i
	+
	\frac{K}{N}
	\sum_{j\in\mathcal{M}}
	A_{ij}(\theta_j-\theta_i)
	-
	\frac{K}{N}
	q_i\theta_i .
\end{equation}
This expression defines a linear dynamics for the non-reset subsystem, where the effect of resetting appears through the local pinning contribution proportional to $q_i$. In order to analyze the collective properties of the system, it is convenient to rewrite the dynamics in matrix form. We introduce the phase vector
\begin{equation}
	\bm{\theta}
	=
	\left(
	\theta_{i_1},
	\theta_{i_2},
	\ldots,
	\theta_{i_M}
	\right)^T,
	\qquad
	i_1,i_2,\ldots {i_M} \in\mathcal M.
\end{equation}
together with the vector of natural frequencies
\begin{equation}
	\bm{\omega}
	=
	(\omega_{i_1},\omega_{i_2},\ldots,\omega_{i_M})^T \qquad
	i_1,i_2,\ldots {i_M} \in\mathcal M.
\end{equation}
Let $\mathbf{A}_{\mathrm{NR}}$ denote the adjacency matrix associated with the induced non-reset subgraph. Using this matrix, we define the Laplacian matrix
\begin{equation}
	\mathbf{L}_{\mathrm{NR}}
	=
	\mathbf{D}_{\mathrm{NR}}
	-
	\mathbf{A}_{\mathrm{NR}},
\end{equation}
where $\mathbf{D}_{\mathrm{NR}}$ is the diagonal degree matrix of the non-reset subsystem. Additionally, we define the diagonal matrix
\begin{equation}
	\mathbf{Q}
	=
	\mathrm{diag}(q_{i_1},q_{i_2},\ldots,q_{i_M}) \qquad
	i_1,i_2,\ldots {i_M} \in\mathcal M,
\end{equation}
whose entries encode the number of reset neighbors connected to each non-reset node. Using these definitions, the linearized dynamics in Eq. (\ref{KM_linear_non-reset}) can be written as
\begin{equation}
	\dot{\bm{\theta}}
	=
	\bm{\omega}
	-
	\frac{K}{N}
	(\mathbf{L}_{\mathrm{NR}}+\mathbf{Q})
	\bm{\theta}.
\end{equation}
The linearized dynamics reveals that the resetting protocol modifies the standard Laplacian dynamics through the addition of a diagonal pinning operator associated with the reset subsystem. Motivated by this structure, we define the grounded Laplacian
\begin{equation}
	\mathbf{L}^{(g)}
	=
	\mathbf{L}_{\mathrm{NR}}
	+
	\mathbf{Q}.
\end{equation}
Using this notation, the dynamics takes the form
\begin{equation}
	\dot{\bm{\theta}}
	=
	\bm{\omega}
	-
	\frac{K}{N}
	\mathbf{L}^{(g)}
	\bm{\theta}.
\end{equation}
The operator $\mathbf{L}^{(g)}$ incorporates simultaneously the connectivity of the non-reset subsystem and the local forcing induced by the reset oscillators. In particular, the reset nodes act effectively as grounded nodes that continuously pin part of the system to the reset phase. In this sense, the operator $\mathbf{L}^{(g)}$ corresponds to the grounded Laplacian associated with the non-reset subsystem.
\\[2mm]
Grounded Laplacians have been extensively studied in networked dynamical systems involving consensus with stubborn agents and leader--follower dynamics, where a subset of nodes is held fixed while the remaining nodes evolve dynamically \cite{Pirani2014,Pirani2016,Xia2017}. Closely related pinning-control approaches have also been investigated in the context of synchronization of complex networks, where the modification of the Laplacian spectrum induced by externally controlled nodes determines the synchronization properties of the network \cite{He2019}. In the present case, the grounded Laplacian emerges naturally in the infinite resetting-rate limit of the stochastic resetting dynamics. In this limit, the resetting mechanism effectively pins the reset nodes and transforms the synchronization dynamics of the non-reset subsystem into a grounded-Laplacian dynamics governed by $\mathbf{L}^{(g)}$.
\\[2mm]
The stationary state $\bm{\theta}^{\star}$ of the linearized dynamics is obtained from the condition $\dot{\bm{\theta}}=0$, which gives
\begin{equation}\label{Stat_Phase_Lg}
	\bm{\theta}^{\star}
	=
	\frac{N}{K}
	(\mathbf{L}^{(g)})^{-1}
	\bm{\omega}.
\end{equation}
Unlike the standard Laplacian operator of the Kuramoto model, the matrix $\mathbf{L}^{(g)}$ possesses no rotational zero mode whenever at least one reset connection exists. Therefore, resetting regularizes the Laplacian dynamics by shifting the spectrum away from zero and ensuring the invertibility of the effective operator.
\\[2mm]
We now analyze the consequences of this structure for synchronization. In the strongly synchronized regime, the complex exponential entering the definition of the order parameter may be expanded as
\begin{equation}
	e^{\mathbf{i}\theta_j}
	\approx
	1+\mathbf{i}\theta_j-\frac{\theta_j^2}{2}.
\end{equation}
Since the synchronization order parameter is invariant under a global phase shift, its leading correction is determined by the fluctuations of the phases around their average value. Introducing the mean phase
\begin{equation}
	\bar{\theta}
	=
	\frac{1}{M}
	\sum_{j\in\mathcal M}
	\theta_j,
\end{equation}
the stationary value of the order parameter defined in Eq.~(\ref{Order_p_non-reset_M}), denoted by $r^{(\mathrm{st})}_{\mathrm{NR}}$, satisfies
\begin{equation}
	r^{(\mathrm{st})}_{\mathrm{NR}}
	\approx
	1
	-
	\frac{1}{2M}
	\sum_{j\in\mathcal M}
	(\theta_j^{\star}-\bar{\theta}^{\star})^2.
\end{equation}
where $\bm{\theta}^{\star}$ denotes the stationary solution of the linearized dynamics given by Eq. (\ref{Stat_Phase_Lg}). This expression shows explicitly that only phase fluctuations relative to the collective phase reduce the synchronization level.
\\[2mm]
The quadratic form above can be written compactly by introducing the projector onto the subspace orthogonal to the uniform vector,
\begin{equation}
	\mathbf P
	=
	\mathbf I
	-
	\frac{1}{M}
	\bm 1
	\bm 1^{T},
\end{equation}
where $\mathbf{I}$ denotes the identity matrix and $\bm{1}$ is the vector whose components are all equal to unity. In terms of $\mathbf{P}$, 
\begin{equation}
	\sum_{j\in\mathcal M}
	(\theta_j-\bar{\theta})^2
	=
	\bm{\theta}^{T}
	\mathbf P
	\bm{\theta}.
\end{equation}
The synchronization order parameter therefore becomes
\begin{equation}
	r^{(\mathrm{st})}_{\mathrm{NR}}
	\approx
	1
	-
	\frac{1}{2M}
	{\bm{\theta}^{\star}}^{T}
	\mathbf P
	\bm{\theta}^{\star}.
\end{equation}
Substituting the stationary solution obtained in Eq.~(\ref{Stat_Phase_Lg}), we obtain
\begin{equation}\label{Eq_r_NR_projector}
	r^{(\mathrm{st})}_{\mathrm{NR}}
	\approx
	1
	-
	\frac{N^2}{2MK^2}
	\bm{\omega}^{T}
	(\mathbf L^{(g)})^{-1}
	\mathbf P
	(\mathbf L^{(g)})^{-1}
	\bm{\omega}.
\end{equation}
Equation~(\ref{Eq_r_NR_projector}) establishes a direct relationship between synchronization and the grounded Laplacian. Unlike the conventional expression based on the Euclidean norm of the phase vector, Eq.~(\ref{Eq_r_NR_projector}) depends exclusively on phase fluctuations orthogonal to the collective phase mode, thereby preserving the invariance of the synchronization measure under global phase shifts.
\\[2mm]
To express this dependence in spectral form, let $\{|\phi_\alpha\rangle\}$ denote the orthonormal eigenvectors of the grounded Laplacian satisfying
\begin{equation}
	\mathbf L^{(g)}
	|\phi_\alpha\rangle
	=
	\mu_\alpha
	|\phi_\alpha\rangle,
\end{equation}
for $\alpha=1,\ldots,M$, where $\mu_\alpha>0$ are the eigenvalues of $\mathbf L^{(g)}$. Using the spectral decomposition
\begin{equation}
	(\mathbf L^{(g)})^{-1}
	=
	\sum_{\alpha=1}^{M}
	\mu_\alpha^{-1}
	|\phi_\alpha\rangle
	\langle\phi_\alpha|,
\end{equation}
the synchronization order parameter can be written as
\begin{equation}\label{Eq_r_NR_Lg}
	r^{(\mathrm{st})}_{\mathrm{NR}}
	\approx
	1
	-
	\frac{N^2}{2MK^2}
	\sum_{\alpha,\beta=1}^{M}
	\frac{
		\langle\bm{\omega}|\phi_\alpha\rangle
		\langle\phi_\alpha|\mathbf P|\phi_\beta\rangle
		\langle\phi_\beta|\bm{\omega}\rangle
	}{
		\mu_\alpha\mu_\beta
	}.
\end{equation}
Equation~(\ref{Eq_r_NR_Lg}) constitutes the central result of the present linear analysis. It shows that synchronization in the non-reset subsystem is determined jointly by the spectral properties of the grounded Laplacian and by the projection onto the subspace orthogonal to the collective phase mode. Consequently, the synchronization level depends not only on the projection of the natural-frequency vector onto the eigenmodes of $\mathbf L^{(g)}$, but also on how these eigenmodes contribute to the physically relevant phase fluctuations selected by the projector $\mathbf P$. 
\\[2mm]
To illustrate the general structure of the linear theory, we now analyze two representative network topologies for the non-reset subsystem. These examples allow us to obtain explicit analytical expressions for the synchronization order parameter and to clarify the role of the grounded Laplacian in the resetting-induced synchronization mechanism. We first consider the limiting case of a complete graph, for which the spectral properties can be obtained exactly. We then analyze Erd\H{o}s--R\'enyi random networks, where the effect of subsystem resetting can be understood within a mean-field approximation. Together, these examples provide complementary insight into how resetting modifies the synchronization properties of both homogeneous and disordered network topologies.
\subsubsection{Complete-graph}
We first specialize the theory to the case in which the non-reset subsystem forms a complete graph. In this situation, every non-reset oscillator interacts with all other oscillators belonging to the subsystem, and the corresponding Laplacian matrix takes the form
\begin{equation}
	\mathbf{L}_{\mathrm{NR}}
	=
	M\mathbf{I}
	-
	\bm{1}\bm{1}^{T}.
\end{equation}
We further assume that the reset nodes are uniformly connected to the non-reset subsystem. Under this condition, every non-reset node possesses the same number of reset neighbors,
\begin{equation}
	q_i=fN,
\end{equation}
where $f$ denotes the fraction of reset oscillators. Consequently, the pinning matrix reduces to
\begin{equation}
	\mathbf{Q}
	=
	fN\mathbf{I},
\end{equation}
and the grounded Laplacian becomes
\begin{equation}
	\mathbf{L}^{(g)}
	=
	M\mathbf{I}
	-
	\bm{1}\bm{1}^{T}
	+
	fN\mathbf{I}.
\end{equation}
The spectrum of this operator can be obtained exactly. The uniform eigenvector, proportional to $\bm{1}$, possesses eigenvalue
\begin{equation}
	\mu_1
	=
	fN,
\end{equation}
whereas the remaining $M-1$ transverse eigenvectors satisfy
\begin{equation}
	\mu_\alpha
	=
	N,
	\qquad
	\alpha=2,\ldots,M.
\end{equation}
Moreover, since the projector onto the subspace orthogonal to the uniform mode satisfies
\begin{equation}
	\mathbf P
	|\phi_1\rangle
	=
	0,
	\qquad
	\mathbf P
	|\phi_\alpha\rangle
	=
	|\phi_\alpha\rangle,
	\qquad
	\alpha=2,\ldots,M,
\end{equation}
the spectral representation of Eq.~(\ref{Eq_r_NR_Lg}) simplifies considerably. The projector removes the contribution of the collective mode, yielding
\begin{equation}
	r^{(\mathrm{st})}_{\mathrm{NR}}
	\approx
	1
	-
	\frac{1}{2MK^2}
	\sum_{\alpha=2}^{M}
	\left|
	\langle
	\phi_\alpha
	|
	\bm{\omega}
	\rangle
	\right|^2.
\end{equation}
Using the completeness of the orthonormal eigenvectors together with the identity
\begin{equation}
	\sum_{\alpha=2}^{M}
	|\phi_\alpha\rangle
	\langle\phi_\alpha|
	=
	\mathbf P,
\end{equation}
the synchronization order parameter can be written equivalently as
\begin{equation}
	r^{(\mathrm{st})}_{\mathrm{NR}}
	\approx
	1
	-
	\frac{
		\bm{\omega}^{T}
		\mathbf P
		\bm{\omega}
	}{
		2MK^2
	}.
\end{equation}
The quantity
\begin{equation}
	\bm{\omega}^{T}
	\mathbf P
	\bm{\omega}
	=
	\sum_{j\in\mathcal M}
	(\omega_j-\bar{\omega})^2,
\end{equation}
where
\begin{equation}
	\bar{\omega}
	=
	\frac{1}{M}
	\sum_{j\in\mathcal M}
	\omega_j,
\end{equation}
is the squared norm of the frequency fluctuations around their mean value. Thus, only the heterogeneous component of the natural-frequency vector contributes to the loss of synchronization, whereas the uniform component is completely removed by the projector. 
\subsubsection{Erd\H{o}s--R\'enyi networks}
We now consider the case in which the dynamics takes place on an Erd\H{o}s--R\'enyi random graph \cite{ErdosRenyi1959}. In this structure, each pair of nodes is connected independently with probability $p$. The network is therefore characterized by a homogeneous connection probability. For $N\gg 1$, the average degree is given by $\langle k\rangle
	\approx
	pN$. Assuming that reset nodes are distributed uniformly at random, the average number of reset neighbors connected to a non-reset node is
\begin{equation}
	\langle q_i\rangle
	=
	pfN.
\end{equation}
Within a mean-field approximation, the pinning matrix may therefore be approximated as
\begin{equation}
	\mathbf Q
	\approx
	pfN\mathbf I,
\end{equation}
so that the grounded Laplacian becomes
\begin{equation}
	\mathbf L^{(g)}
	\approx
	\mathbf L_{\mathrm{NR}}
	+
	pfN\mathbf I.
\end{equation}
The principal effect of subsystem resetting is therefore to shift the Laplacian spectrum according to
\begin{equation}
	\mu_\alpha
	=
	\lambda_\alpha
	+
	pfN,
\end{equation}
where $\lambda_\alpha$ denote the eigenvalues of the Laplacian $\mathbf{L}_{\mathrm{NR}}$ associated with the non-reset Erd\H{o}s-R\'enyi graph.
\\[2mm]
Unlike the complete graph, the projector $\mathbf P$ and the grounded Laplacian $\mathbf L^{(g)}$ do not commute exactly for a generic Erd\H{o}s-R\'enyi network. Consequently, the exact expression for the synchronization order parameter is given by the double spectral sum in Eq.~(\ref{Eq_r_NR_Lg}). The following expression is therefore obtained within the same mean-field approximation used to derive Eq.~(\ref{Eq_r_NR_Lg}), namely that the nontrivial eigenvectors of $\mathbf L^{(g)}$ remain approximately orthogonal to the uniform mode. Under this approximation,
\begin{equation}
	\langle
	\phi_\alpha
	|
	\mathbf P
	|
	\phi_\beta
	\rangle
	\approx
	\delta_{\alpha\beta},
	\qquad
	\alpha,\beta\ge2,
\end{equation}
whereas the uniform mode is annihilated by the projector. The synchronization order parameter then reduces to
\begin{equation}
	r^{(\mathrm{st})}_{\mathrm{NR}}
	\approx
	1
	-
	\frac{N^2}{2MK^2}
	\sum_{\alpha=2}^{M}
	\frac{
		|\langle\phi_\alpha|\bm{\omega}\rangle|^2
	}{
		(\lambda_\alpha+pfN)^2
	}.
\end{equation}
This expression shows that subsystem resetting enhances synchronization by shifting the nontrivial Laplacian spectrum toward larger eigenvalues, thereby reducing the contribution of low-lying modes to the stationary phase fluctuations. The uniform mode does not contribute because it is removed exactly by the projector, reflecting the invariance of the synchronization order parameter under global phase shifts. Consequently, increasing either the reset fraction $f$ or the connection probability $p$ increases the effective pinning strength and leads to a higher degree of synchronization.

\subsection{Mean-field theory in the infinite-resetting limit}

The linear theory developed in the previous section describes the strongly synchronized regime. To extend the analysis beyond this limit, we now formulate a mean-field theory for the infinite-resetting limit, $\lambda\rightarrow\infty$. For an Erd\H{o}s-R\'enyi network, we adopt the annealed-network approximation \cite{Coutinho_PRE_2013,Dorogovtsev_Rev_Mod_2008}, replacing the adjacency matrix by its ensemble average,
\begin{equation}
	A_{ij}
	\simeq
	p \qquad \mathrm{for} \qquad i\neq j.
\end{equation}
This approximation becomes increasingly accurate for sufficiently dense random networks and enables an analytical description of the stationary state. Using Eq.~(\ref{Order_p_non-reset_M}) together with $M=(1-f)N$, the interaction term in Eq.~(\ref{Kuramoto_OD_reset}) becomes
\begin{multline}
	\frac{K}{N}
	\sum_{j\in\mathcal M}
	A_{ij}\sin(\theta_j-\theta_i)
	\\
	\approx
	Kp(1-f)\,
	r_{\mathcal M}
	\sin(\psi_{\mathcal M}-\theta_i).
\end{multline}
Likewise, the interaction with the reset subsystem is approximated as
\begin{equation}
	\frac{K}{N}
	\sum_{j\in\mathcal R}
	A_{ij}\sin\theta_i
	\simeq
	Kpf\sin\theta_i.
\end{equation}
The effective dynamics of a non-reset oscillator is therefore
\begin{equation}
	\dot{\theta}_i
	=
	\omega_i
	+
	Kp(1-f)\,
	r_{\mathcal M}
	\sin(\psi_{\mathcal M}-\theta_i)
	-
	Kpf\sin\theta_i,
	\label{MF_eq_2}
\end{equation}
where the first interaction term is the conventional Kuramoto mean field generated by the non-reset subsystem, while the second represents the effective ordering field induced by the reset oscillators, which continuously drives the dynamics toward the reset phase.

\subsubsection{Symmetric frequency distributions}
\label{Sec_MF_symmetric}
For symmetric frequency distributions $g(\omega)$, the stationary synchronized state is expected to align with the reset phase. We therefore focus on stationary solutions satisfying
\begin{equation}
	\psi_{\mathcal M}=0.
\end{equation}
Throughout this subsection, we consider only stationary solutions. Accordingly, we denote the asymptotic value of the order parameter $r_{\mathcal M}(t)$ by
$r^{(\mathrm{st})}_{\mathrm{NR}}$. Under these conditions, Eq.~(\ref{MF_eq_2}) reduces to
\begin{equation}
	\dot{\theta}_i
	=
	\omega_i
	-
	Kp
	\Big[
	f+(1-f)r^{(\mathrm{st})}_{\mathrm{NR}}
	\Big]
	\sin\theta_i .
	\label{MF_eq_3}
\end{equation}
It is convenient to introduce the effective field
\begin{equation}
	H
	=
	Kp
	\Big[
	f+(1-f)r^{(\mathrm{st})}_{\mathrm{NR}}
	\Big].
	\label{MF_field}
\end{equation}
The dynamics then takes the standard form
\begin{equation}
	\dot{\theta}
	=
	\omega
	-
	H\sin\theta.
\end{equation}
Oscillators satisfying
\begin{equation}
	|\omega|
	\le
	H
\end{equation}
become phase locked, with stationary phases
\begin{equation}
	\sin\theta
	=
	\frac{\omega}{H}.
\end{equation}
The remaining oscillators drift around the circle. Following the standard Kuramoto mean-field treatment, we neglect the contribution of the drifting oscillators to the stationary synchronization order parameter. This approximation is expected to be accurate whenever the synchronized population dominates the coherent response, since the frequency distribution $g(\omega)$ is symmetric about its mean value, $\omega_0=0$.
\\[2mm]
Under this approximation, the stationary order parameter satisfies the self-consistency equation
\begin{equation}
	r^{(\mathrm{st})}_{\mathrm{NR}}
	=
	\int_{|\omega|\le H}
	g(\omega)
	\sqrt{
		1-\frac{\omega^2}{H^2}
	}\,d\omega.
	\label{MF_selfconsistency}
\end{equation}
Equations~(\ref{MF_field}) and (\ref{MF_selfconsistency}) constitute the mean-field description of the stationary synchronization properties in the infinite-resetting limit.
\\[2mm]
A key difference with respect to the ordinary Kuramoto model is that the effective field remains finite even when the stationary synchronization order parameter tends to zero. Indeed,
\begin{equation}
	r^{(\mathrm{st})}_{\mathrm{NR}}
	\rightarrow0
	\qquad
	\Longrightarrow
	\qquad
	H
	\rightarrow
	Kpf .
\end{equation}
Therefore, whenever $f>0$, the incoherent state is no longer an exact stationary solution. The reset subsystem continuously injects coherence into the non-reset population and explicitly breaks the rotational symmetry of the standard Kuramoto dynamics.
\\[2mm]
Further analytical insight can be obtained in the weak-field regime,
$H\ll\Delta_\omega$, where $\Delta_\omega$ denotes the characteristic width of the frequency distribution. Expanding Eq.~(\ref{MF_selfconsistency}) for small $H$ yields
\begin{equation}\label{r_MF_weakH}
	r^{(\mathrm{st})}_{\mathrm{NR}}
	\simeq
	\frac{\pi}{2}
	g(0)\,
	H.
\end{equation}
Using Eq.~(\ref{MF_field}), we obtain
\begin{equation}
	r^{(\mathrm{st})}_{\mathrm{NR}}
	\simeq
	\frac{\pi}{2}
	g(0)
	Kp
	\Big[
	f+(1-f)r^{(\mathrm{st})}_{\mathrm{NR}}
	\Big].
\end{equation}
Solving for $r^{(\mathrm{st})}_{\mathrm{NR}}$ gives
\begin{equation}
	r^{(\mathrm{st})}_{\mathrm{NR}}
	=
	\frac{
		\frac{\pi}{2}
		g(0)
		Kpf
	}{
		1-
		\frac{\pi}{2}
		g(0)
		Kp(1-f)
	}.
	\label{MF_smallH}
\end{equation}
Equation~(\ref{MF_smallH}) explicitly shows that subsystem resetting induces a finite stationary synchronization level even in parameter regions where the ordinary Kuramoto model would remain weakly synchronized.
\\[2mm]
In the absence of resetting, $f=0$, Eqs.~(\ref{MF_field}) and
(\ref{MF_selfconsistency}) reduce to the standard mean-field theory
of the Kuramoto model. In this limit, the synchronization threshold is
\cite{Rodrigues2016}
\begin{equation}
	K_c
	=
	\frac{2}
	{\pi g(0)p},
\end{equation}
which is the classical mean-field result for the onset of synchronization
in dense Erd\H{o}s--R\'enyi networks, where $p$ denotes the connection
probability.
\\[2mm]
For any finite reset fraction $f>0$, however, the effective field remains nonzero even when the stationary synchronization order parameter is arbitrarily small. Consequently, the sharp synchronization transition characteristic of the ordinary Kuramoto model is replaced by a smooth crossover between weakly and strongly synchronized regimes. In this sense, subsystem resetting acts as an external ordering field that continuously promotes coherence and removes the genuine critical behavior present in the absence of resetting.
\\[2mm]
The mean-field theory therefore predicts that increasing either the reset fraction $f$ or the network connectivity $p$ enhances the stationary synchronization level through the effective ordering field generated by the pinned oscillators. This mechanism is analogous to the action of an external field in equilibrium critical phenomena, where symmetry breaking smooths the transition and induces a finite order parameter throughout the system.

\subsubsection{Nonzero mean frequency}
\label{Sec_MF_nonzero}
We now generalize the mean-field theory to the case in which the natural-frequency distribution possesses a nonzero mean,
\begin{equation}
	\omega_0
	=
	\int_{-\infty}^{\infty}
	\omega\,g(\omega)\,d\omega
	\neq0.
\end{equation}
Unlike the symmetric case, the non-reset population possesses an intrinsic tendency to rotate collectively in phase space. Consequently, the average phase $\psi_{\mathcal M}$ cannot be assumed \emph{a priori} to coincide with the reset phase. Instead, both the stationary synchronization level and the average phase must be determined self-consistently.
\\[2mm]
Starting from Eq.~(\ref{MF_eq_2}), it is convenient to rewrite the dynamics as
\begin{equation}
	\dot{\theta}
	=
	\omega
	+
	\mathrm{Im}
	\!\left[
	\mathcal H
	e^{-\mathbf{i}\theta}
	\right],
\end{equation}
where the complex effective field is defined by
\begin{equation}
	\mathcal H
	=
	Kp
	\Big[
	f
	+
	(1-f)
	r^{(\mathrm{st})}_{\mathrm{NR}}
	e^{\mathbf{i}\psi_{\mathcal M}}
	\Big].
\end{equation}
Introducing the modulus and phase of the field,
\begin{equation}
	\mathcal H
	=
	He^{\mathbf{i}\Phi},
\end{equation}
the dynamics assumes the standard form
\begin{equation}
	\dot{\theta}
	=
	\omega
	+
	H\sin(\Phi-\theta),
\end{equation}
which is formally identical to the dynamics of a forced Kuramoto oscillator. The reset subsystem therefore acts as an effective coherent forcing whose amplitude and orientation are determined self-consistently by the collective state of the non-reset oscillators.
\\[2mm]
Assuming the existence of a stationary synchronized state, the phase-locked oscillators satisfy
\begin{equation}
	\omega
	=
	H\sin(\theta-\Phi),
\end{equation}
and therefore contribute to the stationary synchronization order parameter only when $|\omega|\le H$. Neglecting the contribution of drifting oscillators, as in the symmetric case, yields the self-consistency equation
\begin{equation}
	r^{(\mathrm{st})}_{\mathrm{NR}}
	=
	\int_{|\omega|\le H}
	g(\omega)
	\sqrt{
		1-\frac{\omega^2}{H^2}
	}
	\,d\omega.
	\label{MF_nonzero_w0_r}
\end{equation}
\begin{figure*}[t!]
	\begin{center}
		\includegraphics*[width=0.85\textwidth]{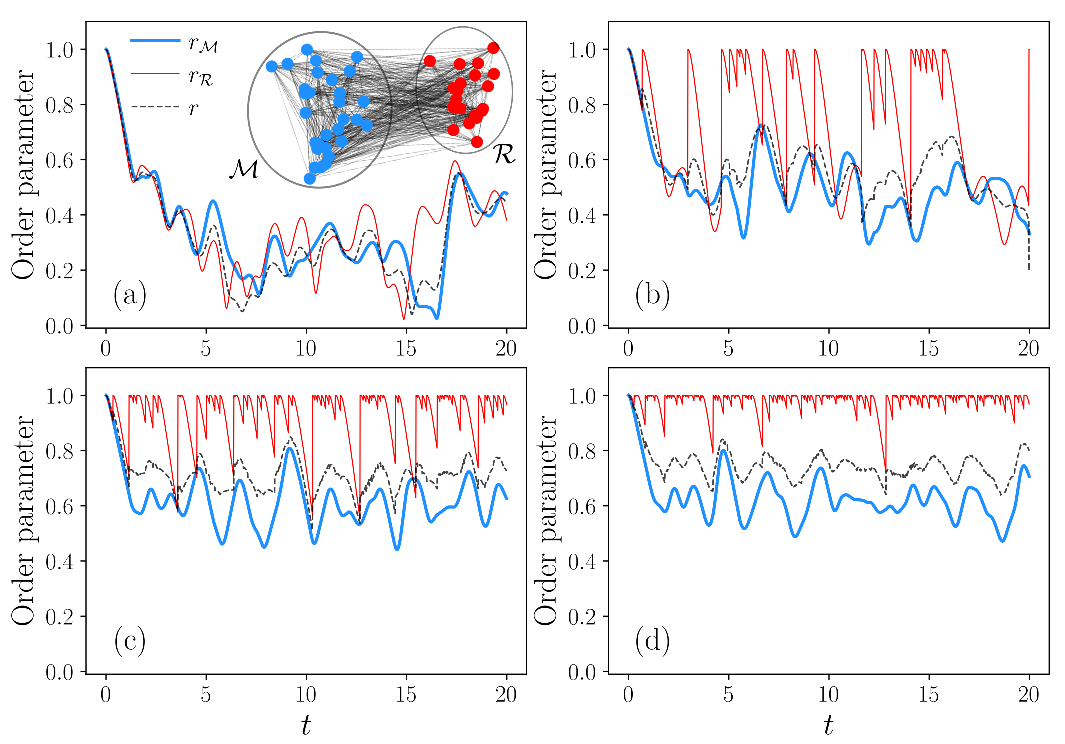}
	\end{center}
	\vspace{-6mm}
	\caption{\label{Fig_1}
		Time evolution of the synchronization order parameters for a network-coupled Kuramoto system with stochastic subsystem resetting. Simulations are performed on a connected Erd\H{o}s--R\'enyi network with $N=50$ oscillators, connection probability $p=0.5$, coupling strength $K=5$, and a fraction $f=0.4$ of oscillators assigned to the reset subsystem. The natural frequencies are independently drawn from a normal distribution with mean $\omega_0=0$ and standard deviation $\sigma=2$. The same realization of the natural frequencies is used in all simulations. Panels (a)--(d) correspond to resetting rates $\lambda=0$, $1$, $5$, and $10$, respectively. The curves show the global synchronization order parameter $r(t)$, the order parameter of the non-reset subsystem $r_{\mathcal{M}}(t)$, and that of the reset subsystem $r_{\mathcal{R}}(t)$. The inset in panel (a) shows the Erd\H{o}s--R\'enyi network together with the partition into the reset subset $\mathcal{R}$ and the non-reset subset $\mathcal{M}$. 
	}
\end{figure*}
Unlike the case $\omega_0=0$, however, the phase
$\psi_{\mathcal M}$ is no longer fixed by symmetry. Instead, it must be determined together with
$r^{(\mathrm{st})}_{\mathrm{NR}}$
through a self-consistent analysis of the collective dynamics. An explicit analytical equation governing the stationary phase generally requires an evolution equation for the complex synchronization order parameter, such as that obtained through the Ott--Antonsen reduction for globally coupled oscillator ensembles \cite{Majumder_PRE_2024,Majumder_PRE2026,Ott_Chaos2008}.
\\[2mm]
The above considerations nevertheless provide a clear physical picture. The nonzero mean frequency induces a collective tendency of the non-reset subsystem to rotate in phase space, while the reset subsystem continuously generates an ordering field directed toward the reset phase. The stationary synchronized state, when it exists, results from the competition between these two mechanisms.
\\[2mm]
If the ordering field dominates the collective drift, the synchronization vector becomes pinned to a fixed orientation relative to the reset phase, leading to a stationary synchronized state characterized by finite values of both
$r^{(\mathrm{st})}_{\mathrm{NR}}$
and
$\psi_{\mathcal M}$.
Conversely, when the collective drift overcomes the restoring action of the reset subsystem, the synchronization vector can no longer remain stationary and instead undergoes persistent rotation. In this regime, the synchronization level remains finite, indicating the persistence of collective coherence, while the average phase evolves continuously in time. The resulting state therefore corresponds to an oscillatory synchronized regime rather than to a stationary synchronized state.

\section{Numerical results}

\subsection{Illustrative dynamics of subsystem resetting}

Before analyzing the stationary properties of the system and dependence of synchronization on the resetting parameters and network topology, it is useful to first examine the elementary dynamical mechanisms generated by subsystem resetting in a finite network. In particular, we are interested in understanding how the repeated resetting of only a fraction of the oscillators influences the collective behavior of oscillators that are never directly reset. Such an analysis provides an intuitive picture of the synchronization process and serves as a guide for the more systematic studies presented in the following sections.
\\[2mm]
Figure~\ref{Fig_1} shows a representative realization of the dynamics on a connected Erd\H{o}s--R\'enyi network composed of $N=50$ oscillators and connection probability $p=0.5$. The inset in Fig. \ref{Fig_1}(a) illustrates the network topology together with its partition into the reset subsystem $\mathcal{R}$ and the non-reset subsystem $\mathcal{M}$. A fraction $f=0.4$ of the oscillators is randomly assigned to the reset subset, while the remaining nodes belong to the non-reset subset. The natural frequencies are independently drawn from the Gaussian distribution with average $\omega_0=0$ and standard deviation $\sigma=2$. The coupling strength is fixed at $K=5$, and the equations of motion are integrated using the Euler method with time step $dt=10^{-2}$ up to a final time $T=20$. Unless otherwise stated, all oscillators are initially prepared in the fully synchronized state $\theta_i(0)=0$. Figures \ref{Fig_1}(a)--(d) correspond to increasing resetting rates, namely $\lambda=0$, $1$, $5$, and $10$, thereby spanning the range from the bare dynamics to a regime of frequent resetting. The three curves shown in each panel correspond to the global synchronization order parameter $r(t)$, the order parameter of the non-reset subsystem $r_{\mathcal M}(t)$, and the order parameter of the reset subsystem $r_{\mathcal R}(t)$. Monitoring these quantities simultaneously allows us to disentangle the direct effect of resetting on the oscillators belonging to $\mathcal{R}$ from the indirect influence exerted on the oscillators in $\mathcal{M}$ through the network interactions.
\\[2mm]
The case $\lambda=0$, shown in Fig.~\ref{Fig_1}(a), corresponds to the ordinary Kuramoto dynamics in the absence of resetting. Since the oscillators possess heterogeneous natural frequencies, the initially synchronized configuration progressively loses coherence and all order parameters decay from their initial value close to unity. A qualitatively different behavior emerges once resetting is introduced. For $\lambda=1$, depicted in Fig. \ref{Fig_1}(b), the order parameter of the reset subsystem remains systematically larger than that of the non-reset subsystem. This behavior reflects the periodic reinjection of coherence produced by the resetting events, which repeatedly drive the oscillators in $\mathcal{R}$ toward the synchronized configuration. More importantly, the enhanced coherence of the reset subsystem is partially transferred to the non-reset oscillators through the network coupling, causing the global order parameter to remain above the values observed in the absence of resetting.
\\[2mm]
As the resetting rate $\lambda$ is increased further as the cases explored in Figs.~\ref{Fig_1}(c) and (d), the synchronization of the reset subsystem becomes increasingly robust. The quantity $r_{\mathcal R}$ remains close to unity during most of the evolution, indicating that the oscillators belonging to $\mathcal{R}$ behave as a highly coherent cluster. At the same time, the order parameter of the non-reset subsystem also exhibits an increase relative to the $\lambda=0$ case, despite the fact that these oscillators are never directly affected by the resetting protocol. This observation constitutes the central phenomenon explored in the present work: resetting a subset of oscillators can induce synchronization in another subset that evolves solely under the bare dynamics.
\\[2mm]
The results displayed in Fig.~\ref{Fig_1} therefore reveal a clear competition between the intrinsic desynchronizing tendency generated by the dispersion of natural frequencies and the coherence continuously injected through the reset subsystem. Increasing the resetting rate strengthens the latter mechanism, leading to higher levels of synchronization both locally and globally. Even in this relatively small network, the figure already provides evidence that subsystem resetting acts as an effective control mechanism capable of promoting collective order through indirect network-mediated interactions.

\subsection{Linear estimate of the required coupling strength}
\begin{figure*}[t!]
	\begin{center}
		\includegraphics*[width=0.99\textwidth]{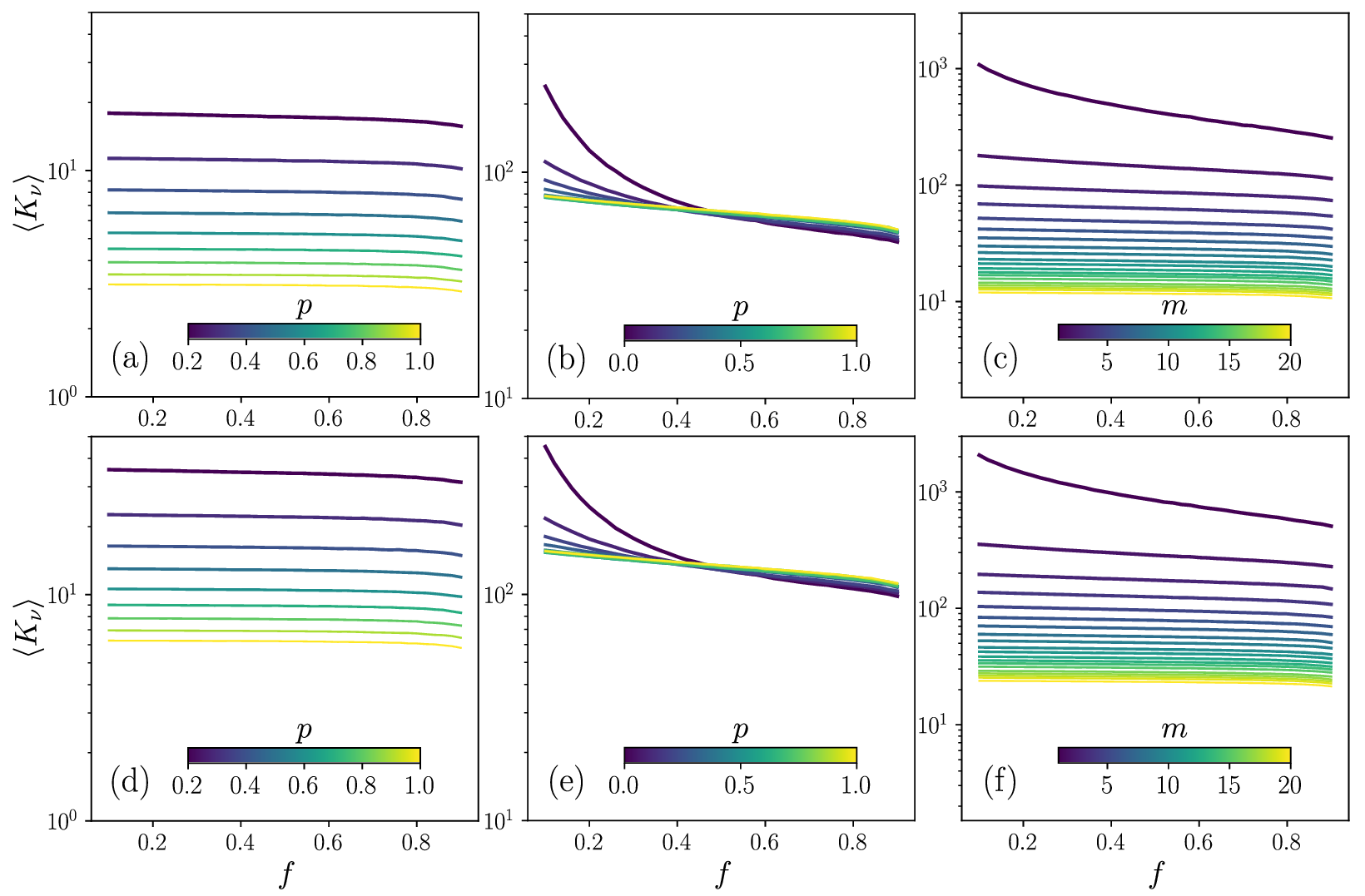}
	\end{center}
	\vspace{-5mm}
	\caption{\label{Fig_2}
		Ensemble average of the coupling strength $\langle K_{\nu}\rangle$ required to achieve the prescribed stationary synchronization level $r_{\mathrm{NR}}^{(\mathrm{st})}=\nu=0.95$ as a function of the reset fraction $f$ for networks of size $N=100$. Panels (a)--(c) correspond to natural frequencies independently drawn from a Gaussian distribution with mean $\omega_{0}=0$ and standard deviation $\sigma=1$, whereas panels (d)--(f) correspond to $\omega_{0}=0$ and $\sigma=2$. Panels (a) and (d) show Erd\H{o}s--R\'enyi networks with link occupation probability $p\in[0.2,1.0]$, as indicated by the color bar. Panels (b) and (e) show Watts--Strogatz networks with initial degree $k=6$ and rewiring probability $p\in[0,1]$, with $p$ indicated by the color bar. Panels (c) and (f) show Barab\'asi--Albert networks generated by preferential attachment, where each newly added node is connected to $m=2,3,\ldots,20$ existing nodes, with $m$ indicated by the color bar. For each network topology, the network realization was kept fixed, while $10^{4}$ independent realizations of the natural frequencies and the reset subset $\mathcal{R}$ were generated to compute the average $\langle K_{\nu}\rangle$ of $K_{\nu}$ using Eq.~(\ref{Eq_K_nu}).		
	}
\end{figure*}
The linear theory developed in Sec. \ref{Sub_Sec_linear_Lg} provides a simple analytical estimate of the coupling strength required to attain a prescribed stationary synchronization level in the non-reset subsystem. Rather than fixing the coupling strength and determining the resulting synchronization level, it is convenient to consider the inverse problem. Specifically, for a prescribed stationary synchronization level $\nu$ close to 1, where the linear approximation is valid, we define $K_{\nu}$ as the coupling strength satisfying
\begin{equation}
	r^{(\mathrm{st})}_{\mathrm{NR}}(K_{\nu})
	=
	\nu.
\end{equation}
Substituting the linear approximation of Eq.~(\ref{Eq_r_NR_projector}) into this condition yields
\begin{equation}
	K_{\nu}
	=
	N
	\sqrt{
		\frac{
			\bm{\omega}^{T}
			(\mathbf L^{(g)})^{-1}
			\mathbf P
			(\mathbf L^{(g)})^{-1}
			\bm{\omega}
		}{
			2M(1-\nu)
		}
	},
	\label{Eq_K_nu}
\end{equation}
which establishes a direct relationship between the coupling strength required to reach the prescribed stationary synchronization level and the grounded Laplacian associated with the chosen reset configuration. Equation~(\ref{Eq_K_nu}) therefore provides a quantitative measure of the effectiveness of subsystem resetting in the limit $\lambda\to \infty$: smaller values of $K_{\nu}$ indicate that the desired stationary synchronization level can be achieved with weaker interactions among the oscillators.
\\[2mm]
\begin{figure*}[t!]
	\begin{center}
		\includegraphics*[width=0.95\textwidth]{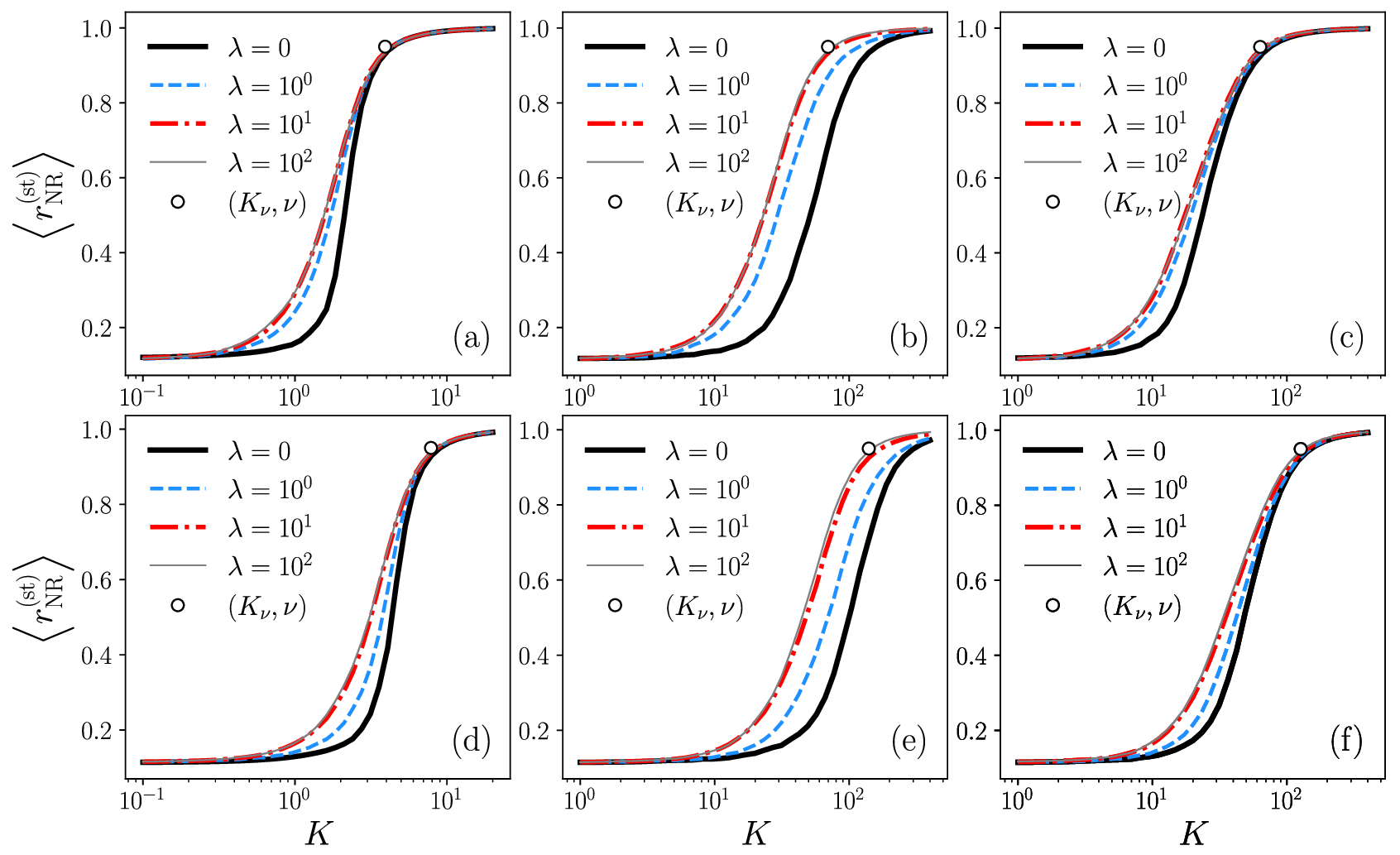}
	\end{center}
	\vspace{-3mm}
	\caption{\label{Fig_3} Ensemble-averaged stationary synchronization order parameter, $\langle r_{\mathrm{NR}}^{(\mathrm{st})}\rangle$, as a function of the coupling strength $K$ for networks of size $N=100$ with a reset fraction $f=0.4$. Panels (a)--(c) correspond to natural frequencies independently drawn from a Gaussian distribution with mean $\omega_{0}=0$ and standard deviation $\sigma=1$, whereas panels (d)--(f) correspond to $\omega_{0}=0$ and $\sigma=2$. The left, middle, and right columns show results for Erd\H{o}s--R\'enyi ($p=0.8$), Watts--Strogatz ($k=6$, rewiring probability $p=0.1$), and Barab\'asi--Albert ($m=4$) networks, respectively. Each curve is averaged over $100$ independent realizations of the natural frequencies while keeping the reset subset $\mathcal{R}$ fixed, and corresponds to a resetting rate $\lambda=0$, $10^{0}$, $10^{1}$, or $10^{2}$. Circle symbols indicate the theoretical values of $K_{\nu}$ predicted by Eq.~(\ref{Eq_K_nu}) for $\nu=0.95$.
	}
\end{figure*}
In Fig.~\ref{Fig_2}, we investigate the dependence of the coupling strength
$K_{\nu}$ required to achieve a prescribed stationary synchronization level
$r^{(\mathrm{st})}=\nu=0.95$ on the reset fraction $f$ for three representative
network models. In addition to Erd\H{o}s--R\'enyi  random graphs, we consider
Watts--Strogatz networks, constructed by rewiring the links of an initially
regular network with probability $p$. As $p$ increases from zero to one, the
network progressively evolves from a regular lattice to a random network, with
a small-world regime emerging for intermediate values of $p$
\cite{WattsStrogatz1998}. We also consider  Barab\'asi--Albert networks,
generated through preferential attachment. In this model, each newly added
node establishes $m$ links to existing nodes with a probability proportional
to their degree, resulting in a heterogeneous degree distribution
\cite{BarabasiAlbert1999}. Figures~\ref{Fig_2}(a)--(c) correspond to natural
frequencies independently drawn from a Gaussian distribution with mean
$\omega_{0}=0$ and standard deviation $\sigma=1$, whereas
Figs.~\ref{Fig_2}(d)--(f) correspond to $\omega_{0}=0$ and $\sigma=2$. For
each network topology, we fix a single connected network realization and
generate $10^{4}$ independent realizations of the natural frequencies and
the reset subset $\mathcal{R}$. The average coupling strength
$\langle K_{\nu}\rangle$ is then obtained from these realizations using
Eq.~(\ref{Eq_K_nu}).
\\[2mm]
Several general features emerge from Fig.~\ref{Fig_2}. First, for both values
of the frequency dispersion and for all three network topologies,
$\langle K_{\nu}\rangle$ generally decreases as the reset fraction $f$
increases. Thus, increasing the number of reset oscillators reduces the
coupling strength required to achieve the prescribed synchronization level.
This result indicates that a larger reset subsystem enhances the propagation
of the synchronizing effect throughout the network. In this sense, the reset
oscillators can be regarded as distributed pinning centers that promote
coherence in the non-reset subsystem. A second feature is the dependence of $\langle K_{\nu}\rangle$ on the amount
of network connectivity. For  Erd\H{o}s--R\'enyi networks [Figs.~\ref{Fig_2}(a) and
\ref{Fig_2}(d)], increasing the link occupation probability $p$ leads to a
systematic reduction in $\langle K_{\nu}\rangle$. Likewise, for Barab\'asi--Albert networks
[Figs.~\ref{Fig_2}(c) and \ref{Fig_2}(f)], increasing the attachment
parameter $m$ decreases the coupling strength required to reach the prescribed
synchronization level. In these two network models, the increase in $p$ or
$m$ is accompanied by an increase in the number of links and, consequently,
in the average degree of the network. The resulting additional connections
provide more pathways through which the synchronizing influence of the reset
subsystem can propagate, thereby reducing the coupling strength required to
maintain the target level of synchronization. 
\\[2mm]
The Watts--Strogatz networks exhibit a qualitatively different behavior
[Figs.~\ref{Fig_2}(b) and \ref{Fig_2}(e)]. In this case, varying the rewiring
probability $p$ does not change the number of links; instead,
it modifies their organization, progressively replacing local connections
with long-range ones. Consequently, the dependence of $\langle K_{\nu}\rangle$
on $p$ cannot be attributed simply to an increase in the number of edges.
Rather, it reflects the influence of network topology on the propagation of
the synchronizing effect generated by the reset subsystem. This dependence
is particularly pronounced for small values of the rewiring probability,
with the largest differences between the curves occurring approximately for
$0\leq p\lesssim0.3$ and $f\lesssim0.4$. As $f$ increases, the curves
corresponding to different values of $p$ become progressively closer,
suggesting that a sufficiently large reset subsystem reduces the sensitivity
of the required coupling strength to the underlying network organization.
\\[2mm]
Finally, comparison of the upper and lower rows of Fig.~\ref{Fig_2} shows
that increasing the frequency dispersion from $\sigma=1$ to $\sigma=2$
increases the coupling strength required to attain the prescribed
synchronization level, as expected. Importantly, however, the qualitative
dependence on the reset fraction and on the network structure is preserved.
Thus, the effects of resetting and network topology observed here are robust
against an increase in the heterogeneity of the natural frequencies.
\\[2mm]
Overall, these results support the theoretical picture emerging from the linear analysis. The coupling strength required to reach a prescribed stationary synchronization level is governed by the grounded Laplacian, which encapsulates the combined effects of the network topology and the reset configuration. More importantly, the decrease of $K_{\nu}$ with increasing reset fraction demonstrates that subsystem resetting provides an efficient mechanism for promoting synchronization even when only a fraction of the oscillators is directly reset.
\subsection{Stationary order parameter as a function of coupling}
The previous subsection focused on the inverse problem of determining the coupling strength required to achieve a prescribed synchronization level. We now consider the complementary problem, namely, the dependence of the stationary synchronization order parameter on the coupling strength. This representation is particularly relevant because the coupling strength is often an intrinsic property of the system, whereas the resetting protocol can be externally controlled. Moreover, it provides a direct way to assess the effectiveness of stochastic resetting in promoting synchronization and to compare the analytical predictions with numerical simulations over a broad range of coupling strengths.
\\[2mm]
Figure~\ref{Fig_3} shows the ensemble-averaged stationary synchronization order parameter, $\langle r_{\mathrm{NR}}^{(\mathrm{st})}\rangle$, as a function of the coupling strength $K$. The results correspond to networks of size $N=100$, with a fraction $f=0.4$ of resetting nodes. The equations of motion are integrated using the Euler method with a time step $dt=10^{-2}$. The stationary order parameter is obtained by numerically integrating the Kuramoto dynamics with stochastic resetting until the transient regime has decayed and subsequently averaging over the stationary portion of the trajectory. Each curve is further averaged over $100$ independent realizations of the natural frequencies while keeping the reset set $\mathcal{R}$ fixed. Four resetting rates are considered: $\lambda=0$, $10^{0}$, $10^{1}$, and $10^{2}$.
\\[2mm]
Figure~\ref{Fig_3} shows the ensemble-averaged stationary synchronization order parameter as a function of the coupling strength for three representative network topologies. Figures \ref{Fig_3}(a)--(c) correspond to natural frequencies independently drawn from a Gaussian distribution with mean $\omega_{0}=0$ and standard deviation $\sigma=1$, whereas Figs. \ref{Fig_3}(d)--(f) correspond to $\omega_{0}=0$ and $\sigma=2$. The left, middle, and right columns show results for Erd\H{o}s--R\'enyi, Watts--Strogatz, and Barab\'asi--Albert networks, respectively. The Erd\H{o}s--R\'enyi networks are generated with link occupation probability $p=0.8$, the Watts--Strogatz networks have initial degree $k=6$ and rewiring probability $p=0.1$, and the Barab\'asi--Albert networks are generated by preferential attachment with $m=4$. The circle symbols indicate the theoretical values of the critical coupling $K_{\nu}$ predicted by Eq.~(\ref{Eq_K_nu}) for the target synchronization level $\nu=0.95$ for $\lambda\gg 1$.
\\[2mm]
For all network topologies considered, Fig.~\ref{Fig_3} shows that increasing the resetting rate shifts the synchronization transition toward smaller coupling strengths, demonstrating that resetting facilitates synchronization. This effect becomes progressively weaker as $\lambda$ increases and is nearly saturated already for $\lambda=10^{1}$. As expected, increasing the frequency heterogeneity from $\sigma=1$ [Figs.  \ref{Fig_3}(a)--(c)] to $\sigma=2$ [Figs.  \ref{Fig_3}(d)--(f)] shifts the synchronization transition to larger values of $K$. The reduction in the synchronization threshold is most pronounced for the Watts--Strogatz networks, whereas Erd\H{o}s--R\'enyi and Barab\'asi--Albert networks exhibit comparatively smaller shifts. In all cases, the theoretical predictions of Eq.~(\ref{Eq_K_nu}), shown as circle symbols, accurately identify the coupling strength at which the stationary synchronization order parameter reaches the prescribed target value $\nu=0.95$.
\subsection{Effects of degree-biased reset node selection}
In complex networks, nodes generally play distinct dynamical roles as a consequence of the heterogeneous connectivity structure. This heterogeneity suggests that the effectiveness of subsystem resetting may depend not only on the fraction of oscillators subjected to resetting but also on how the reset nodes are distributed throughout the network. We therefore consider a biased selection of the reset subset $\mathcal{R}$, in which the probability of selecting node $i$ depends on its degree according to
\begin{equation}
	p_i(\beta)
	=
	\frac{k_i^\beta}{\sum_{l=1}^{N} k_l^\beta},
	\label{p_choice_beta}
\end{equation}
where $k_i$ denotes the degree of node $i$, and $\beta$ controls the selection bias. The reset subset contains $M=fN$ distinct nodes sampled without replacement according to the probabilities in Eq.~(\ref{p_choice_beta}). For $\beta=0$, the reset nodes are selected uniformly, whereas $\beta>0$ favors highly connected nodes and $\beta<0$ favors nodes with smaller degree. Thus, $\beta$ provides a simple control parameter for exploring different strategies for placing the reset nodes while keeping the reset fraction $f$ fixed.
\begin{figure}[t!]
	\begin{center}
		\includegraphics*[width=0.47\textwidth]{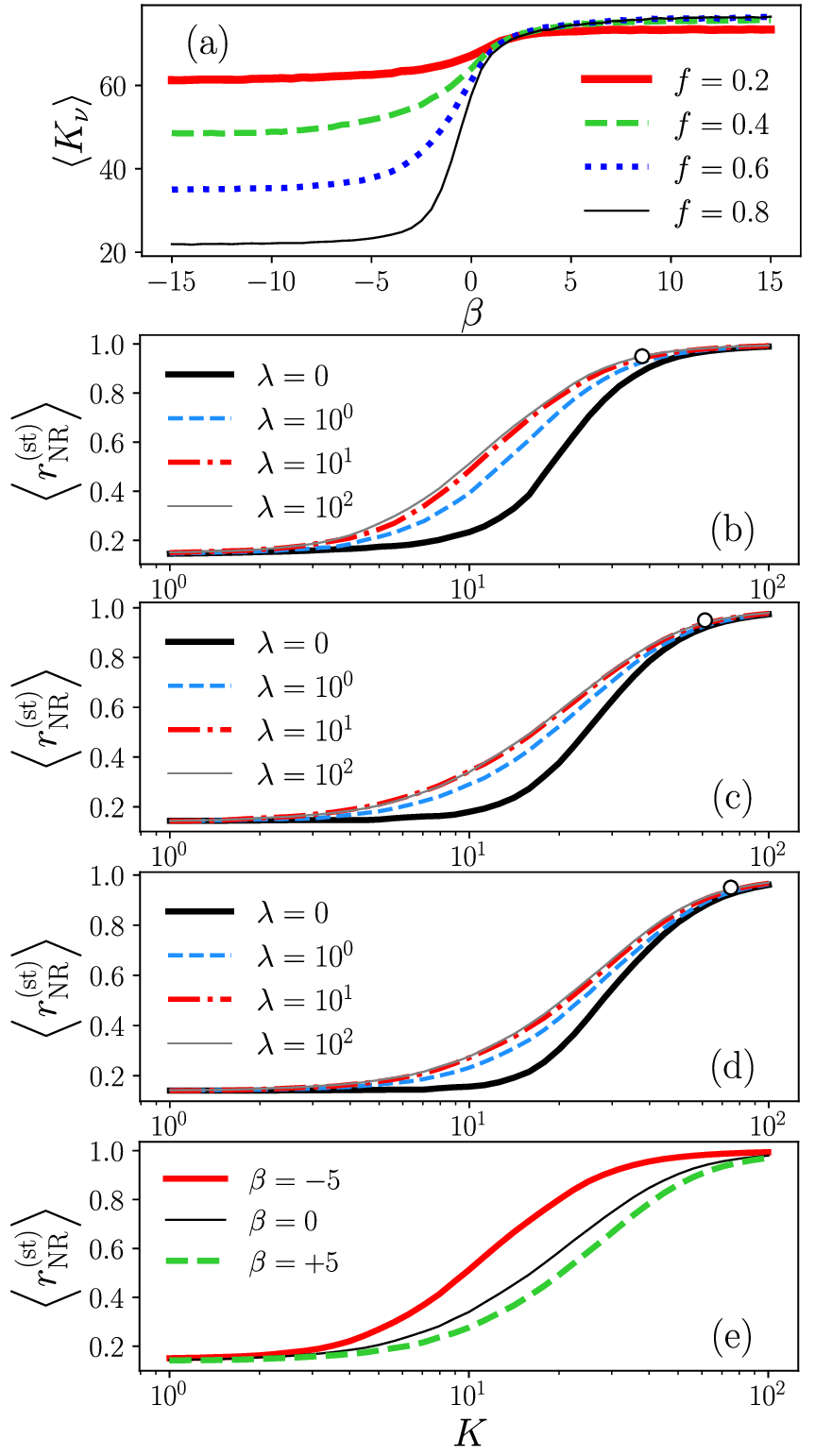}
	\end{center}
	\vspace{-5mm}
	\caption{\label{Fig_4} Effects of biased selection of the reset subsystem $\mathcal{R}$ in a Barab\'asi--Albert network with $N=100$ and $m=4$, with natural frequencies drawn from a Gaussian distribution with mean $\omega_{0}=0$ and standard deviation $\sigma=1$. Panel (a) shows the average $\langle K_{\nu}\rangle$ as a function of the bias parameter $\beta$ for $\nu=0.95$. Panels (b)--(d) show the stationary synchronization order parameter $\langle r_{\mathrm{NR}}^{(\mathrm{st})}\rangle$ as a function of the coupling strength $K$ for a reset fraction $f=0.6$ and different values of the bias parameter: (b) $\beta=-5$, (c) $\beta=0$, and (d) $\beta=5$. Circle symbols indicate the theoretical values of $K_{\nu}$ predicted by Eq.~(\ref{Eq_K_nu}) for $\nu=0.95$. Panel (e) compares $\langle r_{\mathrm{NR}}^{(\mathrm{st})}\rangle$ for $\beta=-5$, $0$, and $5$ at a resetting rate $\lambda=100$.
	}
\end{figure}
\\[2mm]
In Fig.~\ref{Fig_4} we explore the effect of the bias parameter $\beta$ in a Barab\'asi--Albert network with $N=100$ and $m=4$. Its heterogeneous degree distribution provides a natural setting in which the dynamical consequences of biased node selection can be clearly distinguished. The oscillator frequencies at each node are drawn independently from a Gaussian distribution with mean $\omega_{0}=0$ and standard deviation $\sigma=1$. Figure~\ref{Fig_4}(a) shows the average over realizations $\langle K_{\nu} \rangle$ as a function of $\beta$ for a stationary synchronization level $\nu=0.95$. For each value of $\beta$, $10^4$ independent realizations of the natural frequencies and the reset subset $\mathcal{R}$ are generated while keeping the network fixed. The average value $\langle K_{\nu} \rangle$ of $K_{\nu}$ is then calculated using Eq.~(\ref{Eq_K_nu}) for $f=0.2,\,0.4,\,0.6$, and $0.8$. The resulting dependence of $\langle K_{\nu} \rangle$  on $\beta$ demonstrates that the placement of reset nodes can substantially modify the coupling strength required to synchronize the non-reset subsystem. In particular, the increase of $\langle K_{\nu} \rangle$  for positive values of $\beta$ indicates that selecting less connected nodes provides a more effective resetting strategy than selecting highly connected nodes, as the latter requires a larger global coupling strength to achieve the same synchronization level.
\\[2mm]
Figures~\ref{Fig_4}(b)--(d) show the stationary synchronization order parameter $\langle r_{\mathrm{NR}}^{(\mathrm{st})}\rangle$ as a function of the coupling strength $K$ for $\beta=-5$, $0$, and $5$, respectively. The results correspond to a reset fraction $f=0.6$. The equations of motion are integrated using the Euler method with a time step $dt=10^{-2}$. The ensemble averages $\langle r_{\mathrm{NR}}^{(\mathrm{st})}\rangle$ are obtained from 100 curves corresponding to independent realizations, with the natural frequencies and the reset subset $\mathcal{R}$ fixed for each realization. The three cases considered provide representative examples of strong bias toward low-degree nodes, uniform selection, and strong bias toward high-degree nodes. The curves demonstrate that the choice of the reset subset affects the synchronization response even when the total number of reset nodes remains unchanged. For positive $\beta$, the enhanced selection of highly connected nodes shifts the synchronization response toward larger coupling strengths, whereas biasing the selection toward poorly connected nodes is more effective. The circle symbols, obtained from Eq.~(\ref{Eq_K_nu}), provide the corresponding linear estimates of the coupling strength at which the target value $\nu=0.95$ is reached. The contrast between the three biasing strategies becomes particularly transparent in Fig.~\ref{Fig_4}(e), where $\langle r_{\mathrm{NR}}^{(\mathrm{st})}\rangle$ is compared for $\beta=-5$, $0$, and $5$ at a resetting rate $\lambda=100$. For $dt=0.01$, this value of $\lambda$ closely approximates the limit of infinitely fast resetting. The differences between the curves primarily reflect the structural placement of the reset nodes. The ordering of the synchronization levels demonstrates that, in complex networks, selecting reset nodes according to their degree can significantly enhance the effectiveness of subsystem resetting.
\\[2mm]
The combined behavior of $\langle K_{\nu} \rangle$  and $\langle r_{\mathrm{NR}}^{(\mathrm{st})}\rangle$ demonstrates that the efficiency of subsystem resetting depends on the interplay between network topology and the placement of the reset nodes. In particular, for the Barab\'asi--Albert network considered here, favoring low-degree nodes produces a stronger synchronization response than favoring highly connected nodes. These results highlight the importance of the topological organization of the reset subset and suggest that network structure can be exploited to design resetting strategies that achieve a prescribed synchronization level with weaker coupling.
\subsection{Numerical exploration of the mean-field approximation}
\label{sec:MF_numerics}
\begin{figure}[t!]
	\begin{center}
		\includegraphics*[width=0.47\textwidth]{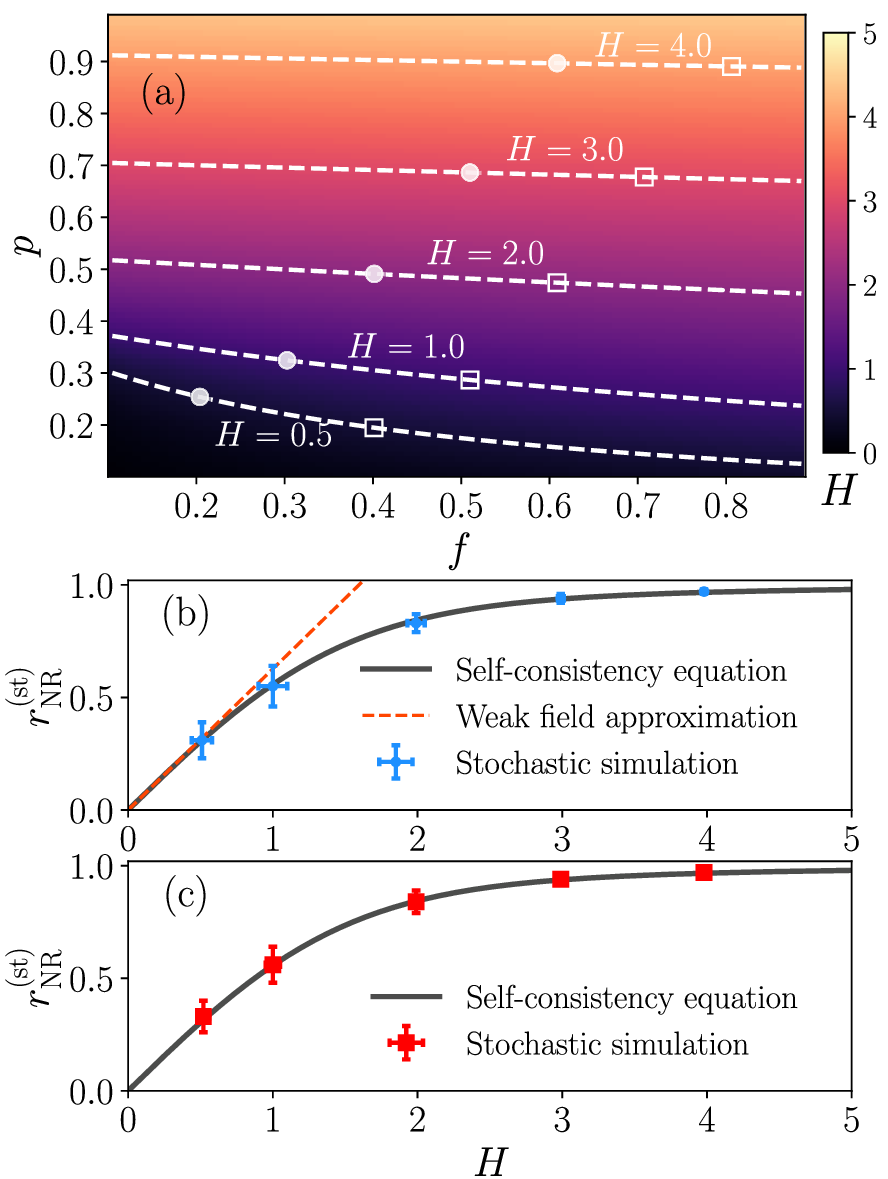}
	\end{center}
	\vspace{-5mm}
	\caption{\label{Fig_5} Mean-field results for Erd\H{o}s--R\'enyi networks with coupling strength $K=4.5$ and a Gaussian frequency distribution with $\omega_0=0$ and $\sigma=1$. (a) Effective field $H$, color-coded in the heatmap, as a function of the reset fraction $f$ and connection probability $p$. Here, $H=Kp[f+(1-f)r_{\mathrm{NR}}^{(\mathrm{st})}]$, with $r_{\mathrm{NR}}^{(\mathrm{st})}$ obtained from Eq.~(\ref{MF_selfconsistency}). Dashed lines indicate constant-$H$ contours for $H=0.5$, $1$, $2$, $3$, and $4$. Circle and square markers indicate the two sets of $(f,p)$ values used in panels (b) and (c), respectively. (b) $r_{\mathrm{NR}}^{(\mathrm{st})}$ as a function of $H$ for the $(f,p)$ values marked by circles in (a). The solid line shows the mean-field prediction from Eq.~(\ref{MF_selfconsistency}), while the dashed line shows the weak-field approximation in Eq. (\ref{r_MF_weakH}). Symbols with error bars show stochastic simulation results averaged over 100 realizations with different frequency assignments. (c) Same as (b), but for the $(f,p)$ values marked by squares in (a).
	}
\end{figure}

Let us now explore numerically the mean-field approximation developed in Sec. \ref{Sec_MF_symmetric}. Having established the self-consistency relation for the stationary state, we test its predictions against direct numerical simulations of the network dynamics. In particular, the mean-field description suggests that, for a symmetric frequency distribution with mean $\omega_0=0$, the effects of the reset fraction $f$ and network connectivity $p$ on the stationary synchronization of the non-reset population enter through the effective field
$H=Kp\left[f+(1-f)r_{\mathrm{NR}}^{(\mathrm{st})}\right]$ in Eq. (\ref{MF_field}). This observation motivates us to use $H$ as the organizing variable for the stationary synchronization state. If the mean-field description is correct, the dependence of the stationary order parameter on $f$ and $p$ should collapse onto a single curve when parametrized by the effective field $H$. In Fig.~\ref{Fig_5}, we test this prediction numerically for a Gaussian frequency distribution with $\omega_0=0$ and $\sigma=1$, considering Kuramoto dynamics on Erd\H{o}s--R\'enyi networks with coupling strength $K=4.5$.
\\[2mm]
\begin{figure*}[t!]
	\begin{center}
		\includegraphics*[width=0.97\textwidth]{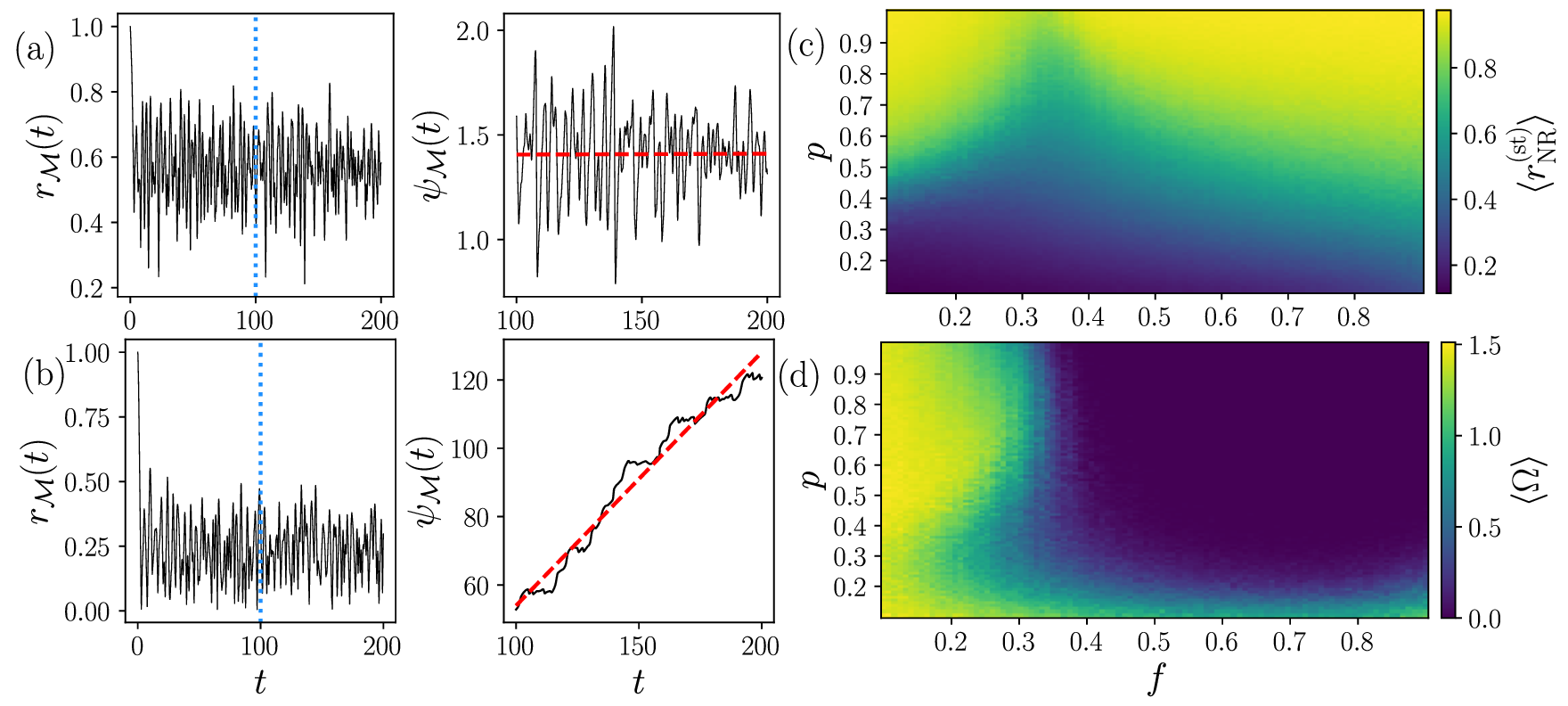}
	\end{center}
	\vspace{-5mm}
	\caption{\label{Fig_6} Numerical characterization of stationary and rotating synchronized states for a Gaussian frequency distribution with mean $\omega_0=1.5$ and standard deviation $\sigma=1$ on Erd\H{o}s--R\'enyi networks with $N=100$ and coupling strength $K=4.5$. (a) Time evolution of the synchronization order parameter $r_{\mathcal M}(t)$ and collective phase $\psi_{\mathcal M}(t)$ for $f=0.5$ and $p=0.6$. In the stationary regime, $100\leq t\leq200$, the collective phase remains approximately constant, as indicated by the dashed horizontal line. (b) Same as (a), but for $f=0.3$ and $p=0.3$. In this case, the synchronization order parameter remains finite, while the collective phase rotates approximately uniformly according to $\psi_{\mathcal M}(t)=\Omega t+\mathrm{const.}$, with $\Omega=0.741$, as indicated by the dashed line. (c) Average stationary synchronization order parameter $\langle r_{\mathrm{NR}}^{(\mathrm{st})}\rangle$ and (d) collective angular velocity $\langle \Omega \rangle$ in the $(f,p)$ parameter plane. The corresponding values are represented by the color scale. The heatmaps in (c) and (d) are obtained from numerical simulations averaged over $100$ independent realizations for each pair $(f,p)$.
	}
\end{figure*}
Figure~\ref{Fig_5}(a) shows the effective field $H$ in the $(f,p)$ parameter plane, with the color scale representing its magnitude. For each pair $(f,p)$, the stationary order parameter $r_{\mathrm{NR}}^{(\mathrm{st})}$ is obtained by solving the self-consistency equation~(\ref{MF_selfconsistency}), after which $H$ is calculated from Eq.~(\ref{MF_field}). The dashed lines correspond to contours of constant effective field, $H=0.5$, $1$, $2$, $3$, and $4$. These contours illustrate that distinct combinations of reset fraction and network connectivity can generate the same effective ordering field. We select two sets of $(f,p)$ values along these contours, indicated by circles and squares, respectively, to test whether the stationary synchronization depends on $f$ and $p$ separately or is determined primarily by their combination through $H$.
\\[2mm]
In Fig.~\ref{Fig_5}(b), we plot $r_{\mathrm{NR}}^{(\mathrm{st})}$ as a function of $H$ for the $(f,p)$ values indicated by circles in Fig.~\ref{Fig_5}(a). The solid curve is obtained by numerically evaluating the mean-field self-consistency relation in Eq.~(\ref{MF_selfconsistency}). The dashed line shows the corresponding weak-field approximation in Eq.~(\ref{r_MF_weakH}). The symbols with error bars represent stochastic simulations of the Kuramoto dynamics on Erd\H{o}s--R\'enyi networks. For each parameter set,  $r_{\mathrm{NR}}^{(\mathrm{st})}$ is obtained by averaging over the measurement interval and subsequently over $100$ realizations with different oscillator frequencies. The corresponding effective field is calculated from the measured time-averaged order parameter according to Eq.~(\ref{MF_field}). The numerical results follow the mean-field prediction over the range of effective fields considered, while the weak-field expression accurately captures the behavior at  $0\leq H\leq 0.5$.
\\[2mm]
Figure~\ref{Fig_5}(c) provides an independent test of the same prediction using the second set of $(f,p)$ values, indicated by squares in Fig.~\ref{Fig_5}(a). Despite the different values of the reset fraction and network connectivity, the simulation results again follow the same mean-field curve obtained from Eq.~(\ref{MF_selfconsistency}). Thus, the two sets of simulations provide consistent results when the stationary order parameter is represented as a function of the effective field $H$ rather than directly as a function of $f$ or $p$.
\\[2mm]
Taken together, these results demonstrate that the stationary synchronization of the non-reset population is well described by the effective-field approach. In particular, the agreement between simulations corresponding to distinct combinations of $f$ and $p$ supports the interpretation of subsystem resetting as an effective symmetry-breaking ordering field. The reset fraction and network connectivity determine the strength of this field through $Kpf$, while the response of the non-reset population is governed by the self-consistent relation between $H$ and $r_{\mathrm{NR}}^{(\mathrm{st})}$. Consequently, within the mean-field approximation, the dependence of the stationary synchronization on the parameters $f$ and $p$ can be reduced to a single response curve $r_{\mathrm{NR}}^{(\mathrm{st})}(H)$. 
\\[2mm]
The mean-field description developed in Sec.~\ref{Sec_MF_nonzero} also shows that, when the frequency distribution has a nonzero mean, synchronization need not correspond to a stationary collective phase. In this case, the non-reset population $\mathcal{M}$ is subject to two competing effects: the nonzero mean frequency favors collective rotation, while the reset subsystem generates an effective ordering field that tends to anchor the collective phase to the reset state. Depending on the relative strength of these effects, the system can therefore exhibit either a phase-locked synchronized state or a rotating synchronized state. In Fig.~\ref{Fig_6}, we investigate this behavior numerically for a Gaussian frequency distribution with $\omega_0=1.5$ and $\sigma=1$ on Erd\H{o}s--R\'enyi networks with $N=100$ and coupling strength $K=4.5$.
\\[2mm]
Figure~\ref{Fig_6}(a) illustrates a phase-locked synchronized state for $f=0.5$ and $p=0.6$. After the transient, for $t\geq 100$, the synchronization order parameter $r_{\mathcal M}(t)$ fluctuates around a finite mean value, $r_{\mathrm{NR}}^{(\mathrm{st})}=0.57$, while the collective phase $\psi_{\mathcal M}(t)$ remains bounded around an approximately constant value, indicated by the horizontal dashed line at $1.4$. Thus, the collective phase is effectively anchored by the reset subsystem, resulting in a stationary synchronized state. In contrast, for $f=0.3$ and $p=0.3$, shown in Fig.~\ref{Fig_6}(b), the order parameter again remains finite after the transient, but the collective phase increases approximately linearly with time. A linear fit in the stationary regime yields
$\psi_{\mathcal M}(t)\simeq\Omega t+\mathrm{const.}$,
indicated by the dashed line, with $\Omega=0.741$. This behavior demonstrates a synchronized state with a persistently rotating collective phase. The finite value of $r_{\mathrm{NR}}^{(\mathrm{st})}$ indicates that the oscillators remain collectively coherent despite the absence of phase locking to the reset state.
\\[2mm]
The occurrence of these two dynamical regimes is further characterized in Figs.~\ref{Fig_6}(c) and \ref{Fig_6}(d). Panel (c) shows the synchronization level averaged over realizations, $\langle r_{\mathrm{NR}}^{(\mathrm{st})} \rangle$, across the $(f,p)$ parameter plane. Panel (d) displays the corresponding average collective angular velocity, $\langle \Omega \rangle$, obtained by linear regression of the collective phase in the stationary regime, following the transient, as illustrated in Fig.~\ref{Fig_6}(b). The values shown in both panels are obtained by averaging over $100$ independent realizations of the oscillator frequencies. The synchronization level increases with network connectivity. In contrast, the collective angular velocity decreases as both the reset fraction $f$ and the network connectivity $p$ increase. For sufficiently strong resetting and network connectivity, $\langle \Omega \rangle$ approaches zero, indicating that the collective phase becomes effectively locked to the reset state. Conversely, for weaker resetting and lower connectivity, $\langle \Omega \rangle$ remains finite, signaling a rotating synchronized regime.
\\[2mm]
These results highlight an important distinction between synchronization and phase locking in the presence of subsystem resetting. A finite stationary order parameter, $\langle r_{\mathrm{NR}}^{(\mathrm{st})}\rangle>0$, characterizes collective coherence but does not, by itself, imply a stationary collective phase. An additional condition, $\langle \Omega\rangle=0$, is required for the synchronization vector to remain locked to the reset phase. Thus, for $\omega_0\neq0$, the resetting field not only affects the degree of synchronization but also competes with the intrinsic tendency of the synchronized population to undergo collective rotation. The numerical results therefore support a picture in which subsystem resetting controls both the degree of collective coherence and the dynamical state of the collective phase. As the resetting strength and network connectivity increase, the system crosses over from a rotating synchronized state with $\langle \Omega\rangle \neq0$ to a phase-locked synchronized state with $\langle \Omega\rangle\simeq0$.
\section*{Conclusions}
We have studied synchronization in network-coupled Kuramoto oscillators subject to stochastic resetting of a subset of nodes. We have shown that resetting can induce synchronization in oscillators that are never directly reset through their coupling to the reset subsystem. In the limit of infinite resetting rate, the reset oscillators become effectively pinned to a reference phase. Linearization of the dynamics near a synchronized state then maps the dynamics of the non-reset subsystem onto a grounded-Laplacian problem. This mapping establishes an analytical connection between the synchronization level and the network structure, frequency heterogeneity, and placement of the reset nodes, and provides an estimate of the coupling strength required to achieve a prescribed degree of synchronization. Beyond the linear regime, we developed a mean-field description in which the reset subsystem acts as an effective ordering field on the non-reset oscillators. For symmetric frequency distributions, this field explicitly breaks the rotational symmetry of the standard Kuramoto dynamics and eliminates the incoherent stationary state for any nonzero fraction of reset oscillators. For frequency distributions with nonzero mean, resetting additionally competes with the intrinsic collective rotation. The resulting synchronized state can therefore be either phase locked to the reset state or characterized by a rotating collective phase, with stronger resetting and network connectivity favoring phase locking. Numerical simulations on different network topologies confirm the predictions of the linear theory in the strongly synchronized regime. The simulations also agree with the mean-field self-consistency relation and demonstrate that the collective response can be characterized in terms of the effective field generated by the reset subsystem.
\\[2mm]
We further find that the effectiveness of subsystem resetting depends not only on the fraction of reset nodes but also on their location within the network. In heterogeneous networks, different reset-node selection strategies can produce substantially different synchronization responses. Thus, subsystem resetting provides a mechanism for inducing and controlling collective synchronization through localized intervention. The grounded-Laplacian and effective-field descriptions provide complementary perspectives on this process, linking the microscopic placement of the reset subsystem to the resulting macroscopic dynamics. Taken together, our results show that local pinning can generate global ordering in networked oscillator systems, even when a fraction of the oscillators is never directly reset, and highlight the role of network structure, dynamical heterogeneity, and reset-node placement in determining the collective response.
\\[2mm]
The framework can be extended to heterogeneous resetting protocols and to directed, weighted, and multilayer networks, as well as to other dynamical processes on networks. Such extensions may provide a basis for designing strategies to control collective dynamics in complex systems through the targeted intervention of a fraction of their agents.

%


%

\end{document}